\documentstyle[aps,preprint,epsf]{revtex}
\def\dir{.}

\draft
\tightenlines
\begin{document}
\title{Randomly Charged Polymers, Random Walks, and Their
Extremal Properties}
\author{Deniz Erta\c s{\dag} and Yacov Kantor{\ddag}}
\address{
{\dag}Department of Physics, Massachusetts Institute of
Technology, Cambridge, MA 02139, U.S.A. \\
{\ddag}School of Physics and Astronomy, Tel Aviv University,
Tel Aviv 69 978, Israel
}
\date{\today}
\maketitle
\begin{abstract}
Motivated by an investigation of ground state properties of
randomly charged polymers, we discuss the size distribution of
the largest $Q$--segments (segments with total charge $Q$) in
such $N$--mers. Upon mapping the charge sequence to
one--dimensional random walks (RWs), this corresponds to finding
the probability for the largest segment with total displacement
$Q$ in an $N$--step RW to have length $L$.
Using analytical, exact enumeration,
and Monte Carlo methods, we reveal the complex structure of
the probability distribution in the large $N$ limit.
In particular, the size of the longest neutral segment
has a distribution with a square-root singularity at
$\ell\equiv L/N=1$, an essential singularity
at $\ell=0$, and a discontinuous derivative at $\ell=1/2$.
The behavior near $\ell=1$ is related to a another
interesting RW problem which we call the ``staircase problem".
We also discuss the  generalized problem for
$d$--dimensional RWs.
\end{abstract}

\section{Introduction}
The importance of understanding proteins\cite{proteins} has attracted
much attention to the statistical mechanics of heterogeneous
polymers. A particular type of heteropolymers built with
a random mixture of positively and negatively charged groups
along their backbone are called polyampholytes (PAs). The
presence of long range electrostatic interactions causes
a rather unique behavior in such polymers: the behavior
of a single PA with unscreened electrostatic interactions
at a low temperature $T$ is extremely sensitive to its
total (excess) charge $Q_o$. Geometrical properties of polymers
can be conveniently described by their radius of gyration
(root--mean--squared size) $R_g$\cite{degen}. At high
$T$, the effect of electrostatic interactions
is small and $R_g$ is approximately equal to that of an
uncharged polymer. However, upon lowering of $T$ the PA
attempts to take
advantage of the presence of two types of charges along its
backbone by assuming spatial conformations in which every
charge is predominantly surrounded by charges of an opposite
sign. This behavior can be approximately described using a
Debye--H\"uckel--type theory\cite{HJ}, which leads to the
conclusion that at low $T$ the polymer should collapse
into an dense state with condensation energy
$E_{\rm cond}\sim-Nq_o^2/a$,
where $N$ is the number monomers, $q_o$ is the typical
charge of a monomer, and $a$ is a microscopic distance such as
diameter of the monomer. In such a collapsed state,
$R_g\sim N^{1/3}$.
On the other hand, renormalization group inspired
scaling arguments showed\cite{KKrg} that at low $T$ one should
expect a strongly stretched state with $R_g\sim N$. This apparent
contradiction was resolved by noting\cite{KKL} that the low--$T$
behavior is extremely sensitive to the overall charge $Q_o$:
It has been observed\cite{KKL} that randomly charged PAs
with {\it vanishing} $Q_o$ indeed collapse at low $T$, while
$R_g$, which is averaged over {\it unrestricted} quenches,
grows with decreasing $T$.
Such sensitivity is consistent with experimental
observations of PAs\cite{experiments}.

\begin{figure}
\centerline{
\epsfxsize=5.9truein
\epsffile{\dir/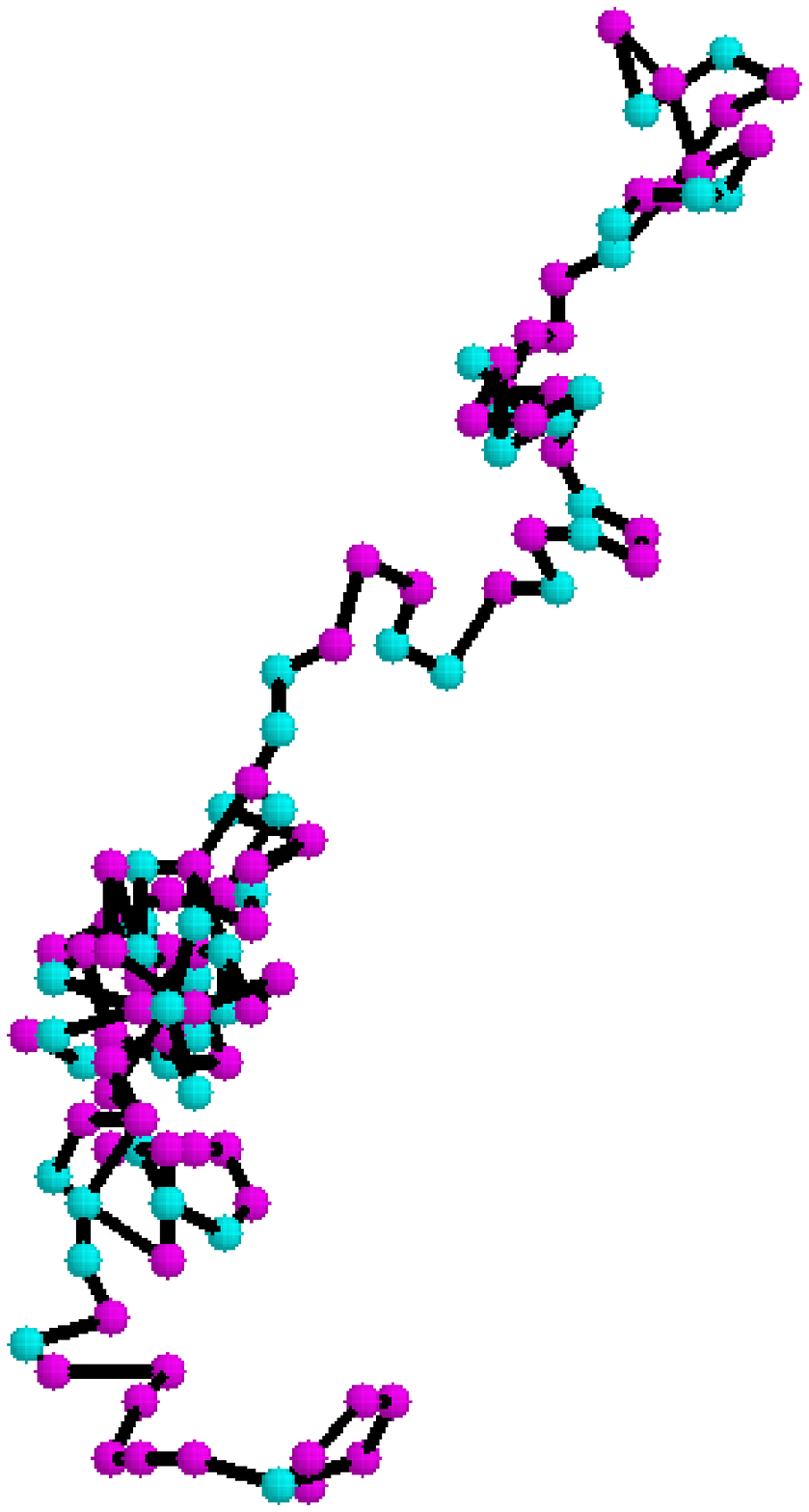}
}
\caption{Low--$T$ configuration of a polyampholyte, which resembles
a necklace made up of weakly charged beads and a highly charged
string.}
\label{PAconf}
\end{figure}

{}From a detailed study of the $Q_o$--dependence of $R_g$, the
following picture began emerging\cite{KK,KKenum}: Consider
a dense (globular, approximately spherical) low--$T$ state
of the PA. Its energy can be roughly separated into
three terms, as
\begin{equation}
E=-N{q_o^2\over a}+\gamma S+{Q_o^2/R_g}\ .
\end{equation}
(In this description we omit the dimensionless prefactors of
order unity.) The first term in this equation represents the
Debye--H\"uckel--type condensation energy, the second term is
the surface energy (where the surface tension
$\gamma\approx q_o^2/a^3$, and the surface area $S\approx
a^2N^{2/3}$), while the last term is the electrostatic
energy of the globule of radius $R_g\approx aN^{1/3}$.
For vanishing $Q_o$, the globule remains approximately
spherical. However, when $Q_o>Q_R\approx q_oN^{1/2}$,
the electrostatic term exceeds the surface tension term,
the spherical shape becomes unstable and the polymer
starts to stretch in order to minimize the electrostatic
energy. Since the  threshold charge $Q_R$
increases with $N$ exactly as the standard deviation of the
total charge $Q_o$ in a random sequence of charges,
for any $N$ there will be a finite portion of chains
with $Q_o$ exceeding $Q_R$. (Note that this
property is specific to three--dimensional electrostatic
interactions. For the $N$--dependence of $Q_R$ in general space
dimensions, see Ref.~\cite{KKenum}.) While the above arguments
suggest that a typical PA should stretch out at low $T$,
such stretching may lead to a loss of the condensation
energy. A reasonable compromise
between stretching (which minimizes the electrostatic energy)
and remaining compact (which gains in condensation energy)
is for the PA to form a {\it necklace} of weakly charged blobs
connected with highly charged ``necks", by taking
advantage of the charge fluctuations along the chain. The
results of the Monte Carlo\cite{KK} and exact
enumeration\cite{KKenum} studies qualitatively
support such a picture. An example of such a low energy
configuration is shown in Fig.~\ref{PAconf}.

While the exact treatment of electrostatic interactions
is not possible, we can pose a simplified problem which,
we hope, captures some essential features of this necklace
model. For example, we may ask what the typical
size of the largest neutral (or weakly charged) segment
in a random sequence of $N$ charges will be.
In order to answer this question, we investigated
the size distribution
of the largest $Q-$segments (segments with a total charge $Q$) in
such $N-$mers.  This problem can be mapped to a one-dimensional
random walk (RW): the sequence of charges $\{q_i\}$
($i=1,\dots,N;\  q_i=\pm1$), is mapped into a sequence of
unit steps in the positive or negative directions along an axis.
The sequence of charges with vanishing total charge $Q_o$ now
corresponds to a RW which returns to the origin after $N$ steps,
while a neutral segment inside the sequence of
charges corresponds to a loop inside the RW. Similarly,
a segment with charge $Q$ corresponds to a segment
(in the corresponding RW) whose end
is displaced by $Q$ units from its beginning. The primary
objective of this work is to investigate the probability
$P_N(L,Q)$ that the {\it largest} $Q$--segment in an $N$--step
RW has length $L$.

There is an apparent simplicity of the formulation of the
problem, i.e. it is similar (and related) to the classical
RW problems\cite{rchandra},  such as the problem
of first passage times or the problem of last return to the
starting point, for which probability
distributions can be computed exactly by using the method
of reflections\cite{mathnote}. However, the search for the
{\it longest} segment of the RW,
among all possible starting points, creates a more
complicated problem.

Some of the results presented in this paper have been briefly
reported before\cite{KEloop}. In this work we present
a complete exposition of those results, as well as many
new results related to this problem and its generalized
version.
In Section \ref{secdefin} we define the problem accurately
and argue that in the large $N$ limit it can be described in
terms of a probability density $p(\ell,q)$, where
$\ell\equiv L/N$ and $q\equiv Q/\sqrt{N}$ are the reduced
length and charge, respectively. This probability density is
investigated using Monte Carlo (MC) and
exact enumeration methods, as well as by analytical arguments.
In particular, we show that the function $p(\ell,0)$ has an
essential singularity in the $\ell\to0$ limit, and diverges
as $1/\sqrt{(1-\ell)}$ in the limit $\ell\to1$.
These properties can be easily understood from qualitative
arguments presented in Section~\ref{secapprox}.
In Section \ref{analytic} we construct an exact
integral expression which enables an analytic investigation of
certain properties of $p(\ell,q)$.
In Section \ref{related} we show that our problem is related
to a different problem of two random walkers (which
we call the ``staircase problem'').
This relation enables us to use the latter problem
to investigate the behavior of $p(\ell,0)$ in the limit
$\ell\to1$. While some of the properties of $p(\ell,q)$ can be
deduced analytically, we had to complement our results by
MC and exact enumeration studies, which
appear in almost every section of the paper along with
analytical arguments on the subject.

An additional insight into the problem can be gained by
considering its generalization to $d$--dimensional RWs.
(This generalization is {\it not} related to the original
problem of PAs or to their embedding dimension.) In this
generalization, which is described in Sec.~\ref{highd}, $\bf Q$
is treated as a $d$--dimensional vector rather than a scalar.
As in the one--dimensional case, ${\bf Q}={\bf 0}$ corresponds
to a loop in the RW. Since the generalized problem investigates
the presence of large loops, it is somewhat related to the
problem of self--avoiding walks\cite{degen}, whose
behavior is also controlled by self--intersections (i.e. loops).
In particular, the probability distribution of the
${\bf Q}$--segments becomes
trivial for $d>4$, when large loops are virtually absent.

\section{Extremal segments: definitions and main properties}
\label{secdefin}

In this Section, we present an exact definition of the
problem of extremal segments of a one dimensional sequence
and review the qualitative features of the resulting
probability distributions.

\begin{figure}
\centerline{
\epsfxsize=5.9truein
\epsffile{\dir/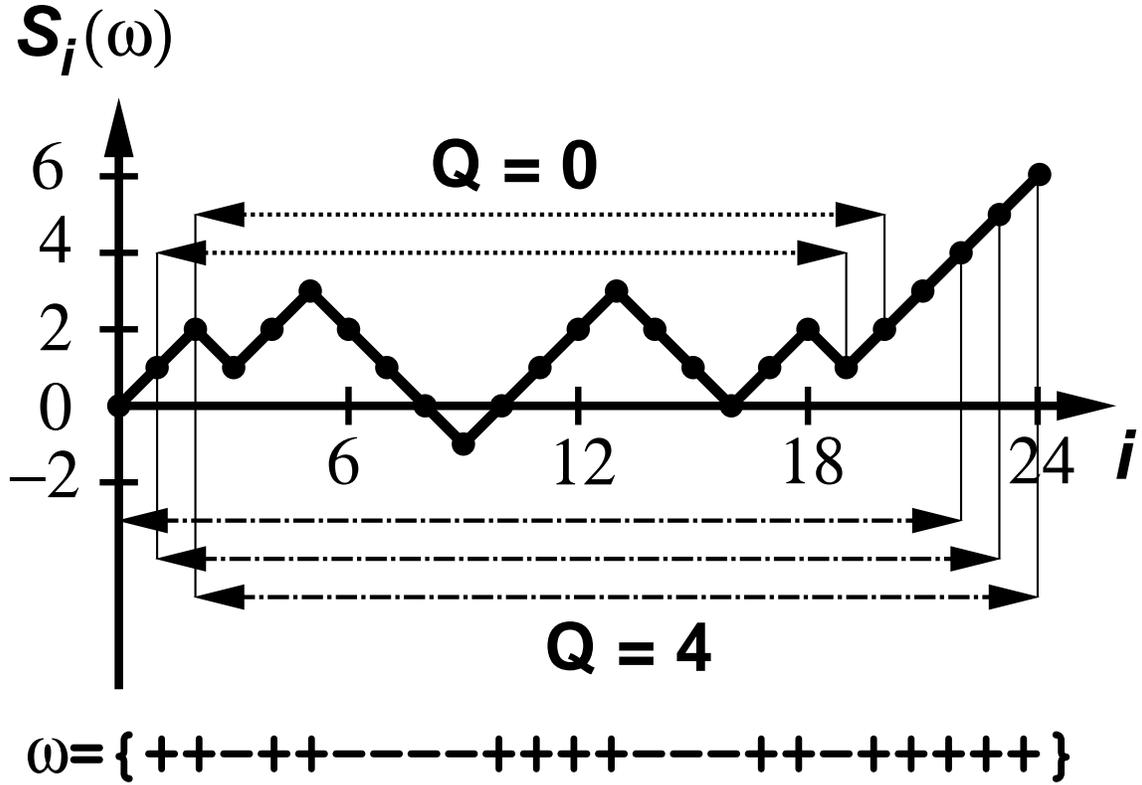}}
\medskip
\caption{Example of a RS $\omega$, and the corresponding RW
depicted by $S_i(\omega)$. In this case, the longest 0--segments
have lengths $L=18$ (dotted lines), while the longest
4--segments (dot--dashed lines) have lengths $L=22$. There are
no 8--segments.}
\label{randomwalk}
\end{figure}

Consider the set $\Omega_N$ which contains all $N$--element
sequences $\{q_i\}$ ($i=1,\dots,N; q_i=\pm1$).
Here, $q_i$ physically corresponds to the charge (positive or
negative) on the $i$th monomer of the $N$-mer. Alternatively,
it can be thought of as the direction of the $i$th step of
an $N$--step one--dimensional RW. A randomly charged polymer
(or, alternatively, RW) can then be represented as a
random sequence (RS) $\omega\in\Omega_N$ picked with equal
probability $2^{-N}$.
Fig.~\ref{randomwalk} depicts an example
of such a sequence and the corresponding path, where the
position  $S_i(\omega)=\sum_{j=1}^iq_j$
of the path at index $i$ gives the accumulated charge
from the beginning of the polymer till the $i$th monomer.
($S_0(\omega)\equiv0$.) In the language of the RWs, $S_i$
is simply the displacement of the walk from the origin after
$i$ steps.
Every segment of the sequence
between, say, steps $i$ and $j$, has a certain charge
$Q_{ij}(\omega)=S_j(\omega)-S_i(\omega)$. A segment for which
$Q_{ij}(\omega)=Q$ will be called a $Q$--segment.
Given a randomly chosen sequence $\omega\in\Omega_N$ and a
charge $Q$, let $P_N(L,Q)$ denote the probability
that the {\it largest} $Q$--segment in $\omega$ has
length $L$. It should be stressed that the definition refers
to the largest $Q$--segment among many possible $Q$--segments
with different starting points which may exist in $\omega$.
For example, the dotted lines in Fig.~\ref{randomwalk} indicate
the longest 0--segments ($L=18$) and the dot--dashed lines show
the longest 4--segments
($L=22$) in a sequence with $N=24$\cite{langfoot}.
Clearly, the longest
$Q$--segment does not have to be unique. If there is at
least one $Q$--segment in the sequence then its length $L$
satisfies $0\le L\le N$. From the definitions is clear that
the 0--segment is always present
and therefore $\sum_{L=0}^N P_N(L,0)=1$. However, the set of $Q$--segments
in a given sequence may be empty for $|Q|>0$: For example, the sequence
shown in Fig.~\ref{randomwalk} has no 8--segments. Thus,
$\sum_{L=0}^N P_N(L,Q)<1$ for $|Q|>0$.

Most properties of RSs have simple continuum limits.
We demonstrate this in Sections ~\ref{secapprox}
and \ref{related} by discussing
RW problems that are exactly solvable, and relating them
to the behavior of $P_N(L,Q)$ in certain limits.
Thus, we also expect $P_N(L,Q)$ to approach a similar scaling
form when $N,L,Q\to\infty$, while the reduced length
$\ell\equiv L/N$ and the reduced charge $q\equiv Q/\sqrt{N}$
are kept constant. In this continuum limit, it is more
convenient to work with the {\it probability density}
\begin{equation}\label{eqdefp}
p(\ell,q)\equiv{N\over 2}\left[P_N(L,Q)+P_N(L+1,Q)\right]\ .
\end{equation}
Of course, for small $N$, this definition of $p(\ell,q)$
will still depend on $N$. We expect it to become a function
only of the reduced variables in the $N,L,Q\to\infty$ limit.
Note that at least one term the the square brackets of
Eq.~(\ref{eqdefp}) vanishes since $P_N(L,Q)=0$ for odd
$L+Q$. To prevent even--odd oscillations, we included
two terms in the definition of $p$,
as in the definitions which are used in continuum
limits of discrete RWs.

\begin{figure}
\centerline{
\epsfxsize=5.9truein
\epsffile{\dir/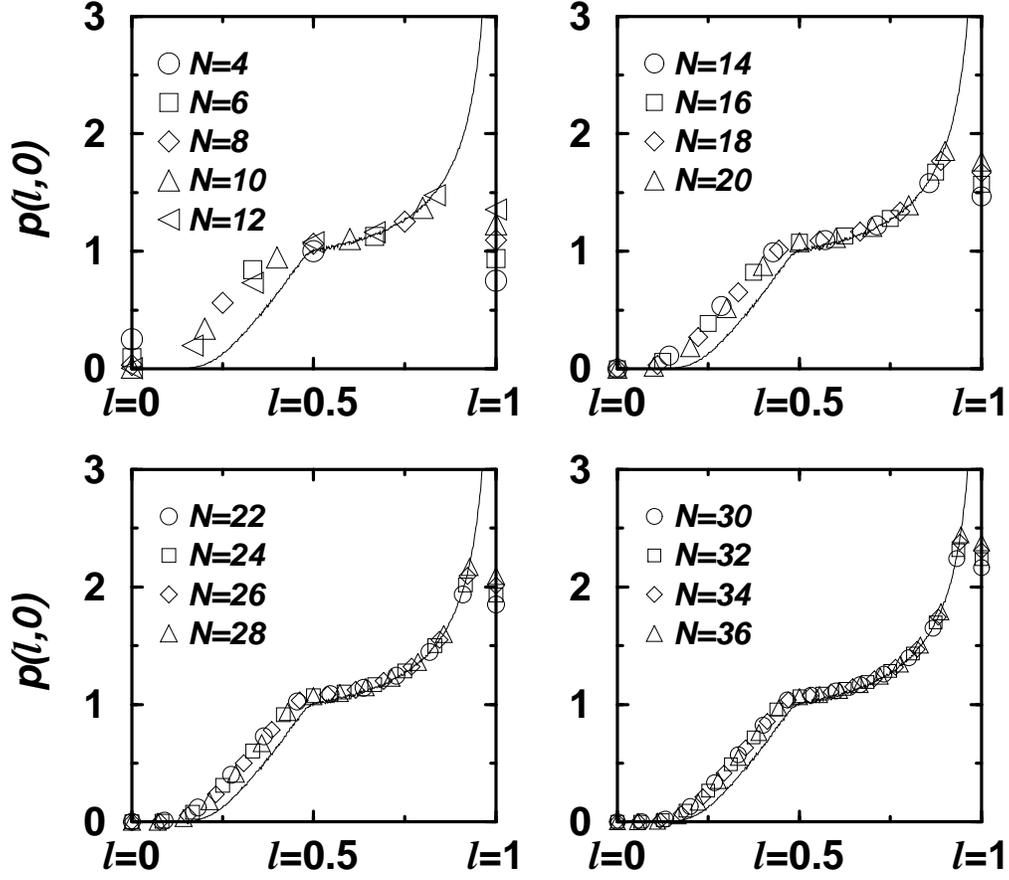}
}
\caption{Probability density of largest neutral segments as a
function of reduced length $\ell=L/N$. Symbols depict
exact enumeration results for $N$ up to 36. In each graph,
the solid line shows
the MC evaluation of $p(\ell,0)$ from $10^8$ randomly
selected sequences of length $N=1000$. }
\label{pl0split}
\end{figure}

We have initially examined the behavior of $P_N(L,Q)$ using numerical
(exact enumeration and Monte Carlo) methods, details of which
are given in the Appendix. Monte Carlo results obtained for
a variety of large $N$s up to $N=10^4$ were virtually
indistinguishable from each other when plotted in the properly scaled
variables. The results for $N=1000$ are depicted as a solid
curve in each one of the graphs in Fig.~\ref{pl0split}. For that
particular value of $N$ we evaluated the probability density
from $10^8$ randomly selected sequences. For short chains
(up to $N=36$) it was possible to perform a complete enumeration
and get the exact results for $P_N(L,Q)$. When these exact
results are plotted in the scaled form, as presented in
Fig.~\ref{pl0split}, we can see that even for such modest
values of $N$, there is an extremely
fast convergence to the continuum distribution $p(\ell,0)$,
depicted by the solid curve (especially for $\ell>0.5$).

The probability density $p(\ell,0)$ shown in Fig.~\ref{pl0split}
has several remarkable properties:

\par\noindent
(a) MC results show
that $p$ at $\ell={1\over2}$ is very close to unity
($1.004\pm0.006$). At that point the slope of the
curve changes by an order of magnitude. While it is impossible
to ascertain from the numerical results that there is actually a
discontinuity in the first derivative of $p(\ell,0)$ with
respect to $\ell$, both the MC results and analytical
arguments indicate that $\ell={1\over 2}$ is a very special
point of the curve.

\par\noindent
(b) For $\ell\to0$, the function exhibits an essential
singularity of the form $\sim \ell^{-x}\exp(-B/\ell)$,
where $B\approx1.7$ and $x\approx 1.5$ -- 2. The estimates
of the coefficient $B$ and of the exponent $x$ have been
obtained from the MC data. However, in the $\ell\to0$ limit
we are dealing with almost vanishing probabilities, and
therefore the statistical accuracy is small. Thus the
estimates depend on the precise range of $\ell$s for which
the fit is performed. Nevertheless, the existence of the
singularity can be easily understood from the fact that for
small $\ell$ the absence of large loops in the entire chain
can be though of as requirement that such loops are absent in
many separate and independent segments of the sequence. In the
Section~\ref{secapprox} this argument will be discussed in
detail.

\par\noindent
(c) For  $\ell\to1$, $p(\ell,0)$
diverges as $A/\sqrt{\pi(1-\ell)}$, with $A=1.008\pm0.005$.
This estimate of the constant $A$ has been obtained from MC results for
$N=1000$ sequence.
In Section~\ref{analytic} we prove the
existence of the square--root singularity from an integral
relation which is derived for $p(\ell,q)$. The proof, however,
does not provide a value for the prefactor $A$, and we are
limited to MC estimates, as well as results extracted
from exact enumeration studies which will be presented in
Sec.~\ref{analytic}. (The accumulated evidence of MC and exact
enumeration shows that $A$ is definitely larger than 1.)
Some more intuitive, although less rigorous,
results regarding the $\ell\to0$ and $\ell\to1$ limits
are presented in Section~\ref{secapprox}. The exact enumeration
results depicted in Fig.~\ref{pl0split} are not suitable for
extraction of asymptotic behavior, since the $N$s are too small.
In the Section~\ref{analytic} we show that it is possible to
exactly calculate $P_N(L=N-M,Q)$ for small $M$ (i.e. $M=0,2,4$) and
{\it arbitrary} sequence length $N$. In principle, the
correct behavior
of $p(\ell,0)$ in the $\ell\to1$ limit can be deduced from the
exact values of $P_N(L=N-M,0)$ only if the limit
$N,M\to\infty$ (while keeping $M/N=1-\ell$ constant) is taken
before the $\ell\to 1$ limit.
Somewhat surprisingly, if we attempt to match
the asymptotic form of $p(\ell,0)$ near
$\ell=1$ with $P_N(L,0)$ for $L=N-2$, we
find $A=1$, i.e. we reproduce almost
the exact value of the prefactor. Thus, the discrete
distribution approaches its asymptotic (continuum) form
within a few steps of the extreme $L=N$.

\begin{figure}
\centerline{
\epsfxsize=5.9truein
\epsffile{\dir/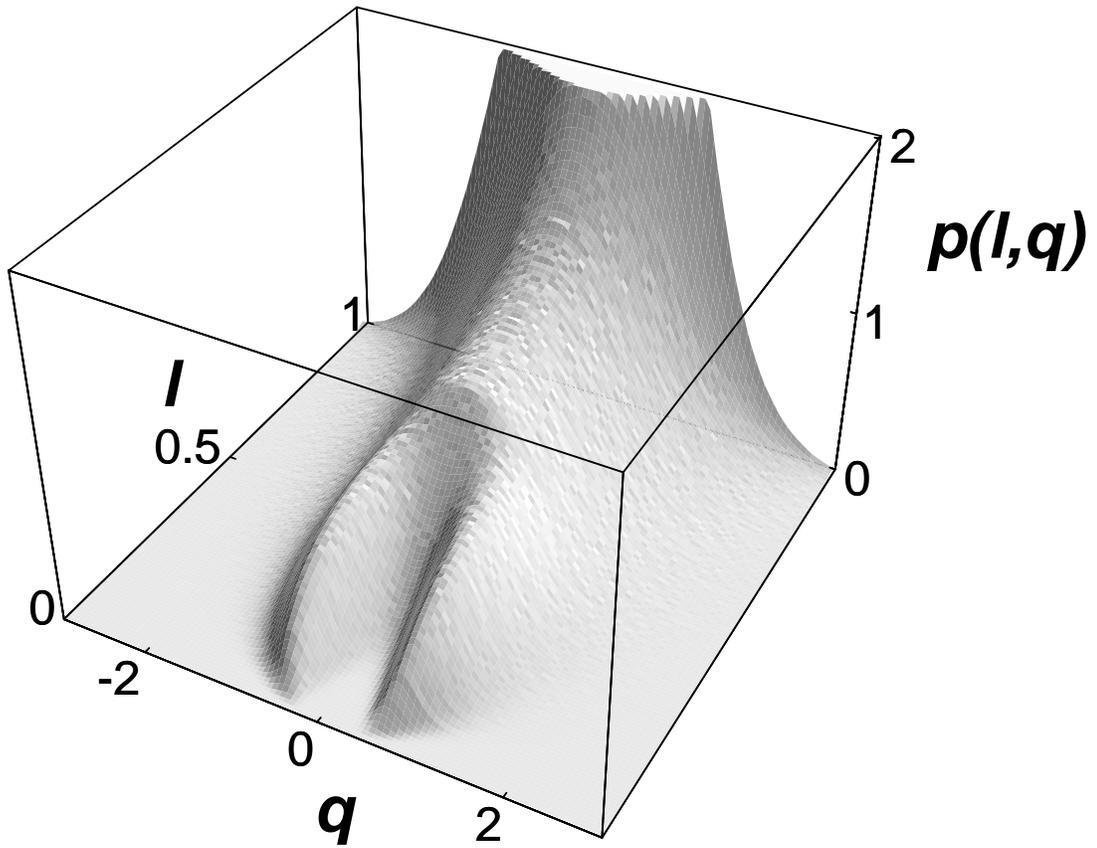}
}
\caption{Probability density of largest $Q$--segments
as a function of reduced charge $q$ and reduced length $\ell$.
The results have been obtained from MC simulations (see text).}
\label{plq}
\end{figure}

Consider next the full probability density $p(\ell,q)$,
which is depicted in Fig.~\ref{plq}. Introduction of an
additional variable $q$ significantly increased the CPU time
needed to analyze a single RS. The MC data in this figure represent
only  $10^7$ sequences of length $N=1024$, i.e. its accuracy
is smaller than the MC results depicted by the solid line in
fig.~\ref{pl0split}.
Fig.~\ref{plq} demonstrates further peculiarities of $p(\ell,q)$:
For fixed $\ell$, the $q$-dependence of $p$ is qualitatively
different for $\ell >{1\over2}$ and $\ell < {1\over 2}$:
When $\ell >{1\over2}$, the distribution has a single
peak at $q=0$, which approaches a Gaussian shape as $\ell$
increases, while for $\ell<{1\over2}$ we see a minimum at
$q=0$ and two peaks symmetrically located around the minimum.
While qualitatively such behavior can be easily understood
(e.g. for small $\ell$ the 0--segments are very unprobable,
since they are typically large, and consequently the maximum
must be reached for non--zero value of $q$) the transition
between the $\ell<{1\over2}$ and $\ell>{1\over2}$ regions is
rather sharp: we analyzed the $q$--dependence of the graphs
representing the fixed--$\ell$ sections of Fig.~\ref{plq}
and concluded that the transition from single maximum to a
minimum surrounded by two maxima cannot be obtained by a
variation of parameters in a simple function (the way it is done
in the mean--field description of a phase transition near the
critical temperature). The numerical data creates an impression
of two different functions glued along $\ell={1\over2}$.

\begin{figure}
\centerline{
\epsfxsize=5.9truein
\epsffile{\dir/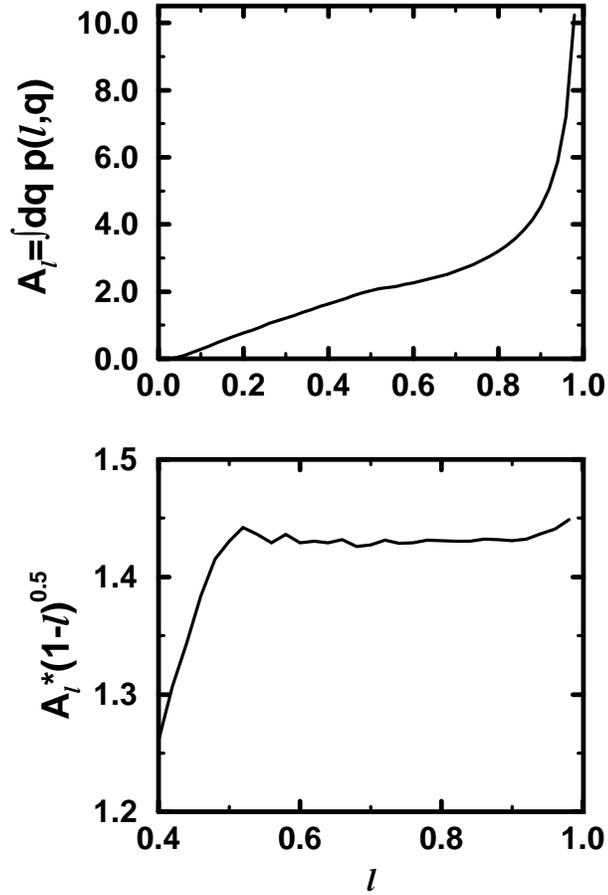}
}
\caption{(a): A plot of the areas $A_\ell$ computed from the
distribution in Fig.~\protect{\ref{plq}}. (b): Demonstration of
the relation $A_\ell\protect\sim1/\protect\sqrt{1-\ell}$
for $\ell>1/2$.
}
\label{intplq}
\end{figure}

The areas
$
A_\ell\equiv\int_{-\infty}^{+\infty}dq\,p(\ell,q)
$
under fixed-$\ell$ sections are shown in Fig.~\ref{intplq}.
For $\ell>{1\over2}$  it will be proven in Sec.~\ref{analytic}
that $A_\ell\sim{\rm const}/\sqrt{1-\ell}$; Fig.~\ref{intplq}b
demonstrates the numerical validity of this relation ---
$A_\ell\sqrt{1-\ell}$ remains approximately constant in the
range of validity. The accuracy of small $\ell$ regime is rather
low; we only note that $A_\ell$
is approximately linear in $\ell$ for $0.15 < \ell < 0.5$, as can be seen
from Fig.~\ref{intplq}a.

\section{Qualitative arguments}\label{secapprox}

In this section we present approximate derivations of several
features of $p(\ell,q)$. Despite the approximate nature of the
arguments, they are rather intuitive, and will be useful when we
generalize the problem to $d$--dimensional RWs.

Most properties of RWs have simple continuum limits. As an
example, let us consider the special case
$L=N$ of our probability distribution:
The probability $P_N(L=N,Q)$ that the largest
$Q$--segment has length $N$ is simply equal to the probability
that the overall charge $Q_o$ of the RS is equal
to $Q$. This probability (for even $N+Q_o$) is given by
\begin{eqnarray}
\label{allcharge}
W_N(Q_o) &\equiv& {\rm Prob}\{S_N(\omega)=Q_o\}
= 2^{-N}\frac{N!}{[(N-Q_o)/2]![(N+Q_o)/2]!} \\
&\displaystyle{=}\atop{N\to\infty}& \quad
\sqrt{\frac{2}{\pi N}}\exp(-Q_o^2/2N). \nonumber
\end{eqnarray}

Consider a restricted subset of all RSs in $\Omega_N$ which
consists only of sequences with {\it total} charge $Q_o$.
The {\it conditional} probability for the largest $Q$--segment
in a sequence selected from this subset to have length $L$
will be denoted as $P_N(L,Q|Q_o)$.
This probability is related to $P_N(L,Q)$ by the relation
\begin{equation}
P_N(L,Q)=\sum_{Q_o} P_N(L,Q|Q_o)W_N(Q_o)\ .
\end{equation}
In the case of
$Q=0$, i.e. for 0--segments, we note that from the definition
it follows that the conditional probability is normalized, i.e.
$\sum_L P_N(L,0|Q_o)=1$. We further note that as a function of
$L$, the conditional probability is expected to be peaked at
value which depends on $Q_o$. Let us assume for
simplicity that the peak is very narrow, i.e. the length of
the largest 0--segment is uniquely determined by $Q_o$ and can
be described by a function $Q_o(L)$.
Indeed, when
$Q_o\approx 0$, the longest 0--segment typically has
$L\approx N$, while for very large $Q_o$, the longest
0--segment must be short. Thus $Q_o(L)$ is a monotonically
decreasing function.
This approximation is especially reasonable for
the extremes $\ell\to0$ or 1. In that case,
$P_N(L,0)\approx W_N(Q_o(L))$,
and thus
\begin{equation}\label{eqapproxp}
p(\ell,0)\approx{N\over 2}W_N\left(Q_o(L)\right)
\left|{dQ_o\over dL}\right|.
\end{equation}
Standard scaling arguments suggest that
for $Q_o\ll\sqrt{N}$ we can relate  $L\approx N-aQ_o^2$,
where $a$ is of order unity. This gives
$Q_o(L)\approx\sqrt{(N-L)/a}$, and finally leads to
\begin{equation}
p(\ell,0)\approx\left({N\over 2}\right)\sqrt{2\over \pi N}
\left({1 \over \sqrt{a(N-L)}}\right)=
\frac{\rm const}{\sqrt{\pi(1-\ell)}}.
\end{equation}
On the other hand, for
$Q_o\gg\sqrt{N}$, the length of the longest 0--segment
will be of order of a scale at which the random excursion of
the RW becomes comparable to the drift produced by $Q_o$,
i.e., when $L^{1/2}\approx (2B)^{-1/2} L Q_o/N$,
where $B$ is a constant of order unity. Thus,
$Q_o(L)\approx N\sqrt{2B/L}$ and
\begin{equation}
p(\ell,0)\approx \left({N\over 2}\right)\left(\sqrt{2\over \pi N}
e^{-BN/L}\right)
\left({\sqrt{B/2}N\over L^{3/2}}\right)=
{{\rm const} \over \ell^{3/2}}e^{-B/\ell}.
\end{equation}
Thus, this simple scaling argument correctly
reproduces the square-root divergence for
$\ell\to1$, and the $\exp({\rm const}/\ell)$ singularity
for $\ell\to0$.

It is useful to consider an alternative derivation of the
behavior in $\ell\to0$ limit, since such derivation involves a
somewhat different view of the same properties.
A RS with an extremely short 0--segment must have a
strong imbalance between the charges (large $Q_o$),
i.e. resemble a biased random walk.
Consider the probability $Z_N(L)=\sum_{L'=0}^L P_N(L',0)$
that the largest 0--segment in an $N$--step sequence {\it does
not exceed} length $L$. If $L\ll N$, this quantity can be
used to estimate $Z_{2N}(L)$ for a sequence twice as long:
Two halves of the sequence of length $2N$ must be biased walks
with the same direction of bias to prevent creation of long
loops, which start in
one half of sequence and end in the other half. In addition,
loops longer than $L$ must be absent from each half of the
sequence. Thus, $Z_{2N}(L)\approx {1\over2}Z_{N}^2(L)$.
This relation is only approximate since it disregards the
correlation between the two halves of the sequence close to
its middle. (Loops longer than $L$ can begin in one half of
the sequence and end at the other half; correction for this
effect may introduce an $L$--dependent prefactor.)
If the continuum limit is well defined, we
can express this relation in the form
\begin{equation}
\int_0^{\ell/2}p(\ell,0)d\ell\approx {1\over2}\left(\int_0^\ell
p(\ell,0)d\ell\right)^2\ .
\end{equation}
This relation is satisfied by $p(\ell,0)=(2B/\ell^2){\em
e}^{-B/\ell}$. The approximation casts serious doubts
on the exact
value of the power of $\ell$ in the prefactor to the
exponential.

Note that two different derivations of the behavior of
$p(\ell,0)$ in the $\ell\to0$ limit produced different
preexponential powers. As was mentioned in the previous Section,
our MC results are not accurate enough to
distinguish between these predictions.

The method of reflections is a standard tool in calculating the
behavior of random walkers near reflecting or absorbing walls
(see Ref.~\cite{rchandra}). It can be used to calculate
various seemingly nontrivial probabilities in terms of
probabilities that are easily evaluated. One such result,
which is important for the following discussion, is that
the probability for an $N$--step RW to never return to its
starting point is equal to the probability that it
reaches its starting point {\it exactly} at the
$N$th step\cite{Feller2}, i.e.,
\begin{equation}
\label{identity}
{\rm Prob}\{S_i(\omega)\neq 0,\ 1\le i \le N\}
={\rm Prob}\{S_N(\omega)=0\}=W_N(0),
\end{equation}
where $W_N$ was defined in Eq.~(\ref{allcharge}).
This relation permits, for instance, an exact solution to a
simplified version of our problem. In the modified problem,
the largest $Q$--segments are selected among those
that {\it start from the beginning} of the RS, rather than all possible
starting positions. This modified probability  $P'_N(L,Q)$ is given by
the probability that the path $\omega$ reaches position $Q$ at the
$L$th step, and that it never again passes through position $Q$
until the $N$th step. Using Eqs.(\ref{allcharge}) and
(\ref{identity}), we obtain the result
(for  $N$, $L$ and $Q$ all even or all odd)
\begin{eqnarray}\label{pprime}
P'_N(L,Q)&=&W_L(Q)W_{N-L}(0)=
2^{-N}\frac{L!}{[(L-Q)/2]![(L+Q)/2]!}
\frac{(N-L)!}{[(N-L)/2]![(N-L)/2]!}\nonumber \\
&\displaystyle{=}\atop{N\to\infty}& \quad
\frac{2}{\pi\sqrt{L(N-L)}}\exp(-Q^2/2L).
\end{eqnarray}
Unfortunately, the search for the longest $Q$--segment in the
RS among all possible starting points creates a more complicated
problem. However, we similarly expect $P'_N(L,Q)$ to approach
a scaling form when $N,L,Q\to\infty$, while the reduced length
$\ell\equiv L/N$ and the reduced charge $q\equiv Q/\sqrt{N}$
are kept constant. In this continuum limit, it the
{\it probability density} is defined analogously with $p$:
$
p'(\ell,q)={N\over 2}[P'_N(L,Q)+P'_N(L+1,Q)].
$
In this limit Eq.~(\ref{pprime}) reduces to
\begin{equation}
p'(\ell,q)=\frac{1}{\pi\sqrt{\ell(1-\ell)}}\exp(-q^2/2\ell).
\end{equation}
We intuitively expect $p$ and $p'$  to behave similarly,
at least in the $\ell\to1$ limit, and indeed in that limit
$p'$ resembles $p$ (see Eq.~(\ref{plimit}))

\section{Exact relations}\label{analytic}

The probabilities $P_N(L,Q)$ for different values of $N$, $L$
and $Q$ satisfy an interesting relation, which in the continuum
limit becomes an integral expression that relates $p(\ell,q)$
at arbitrary values of $\ell>{1\over2}$ and $q$ to the values
of $p(\ell={1\over2},q)$. While such a relation is
insufficient to completely determine the function $p(\ell,q)$,
it suffices to determine some of its important features.
In this Section, we derive this relation and explore
its consequences.

\begin{figure}
\centerline{
\epsfxsize=5.9truein
\epsffile{\dir/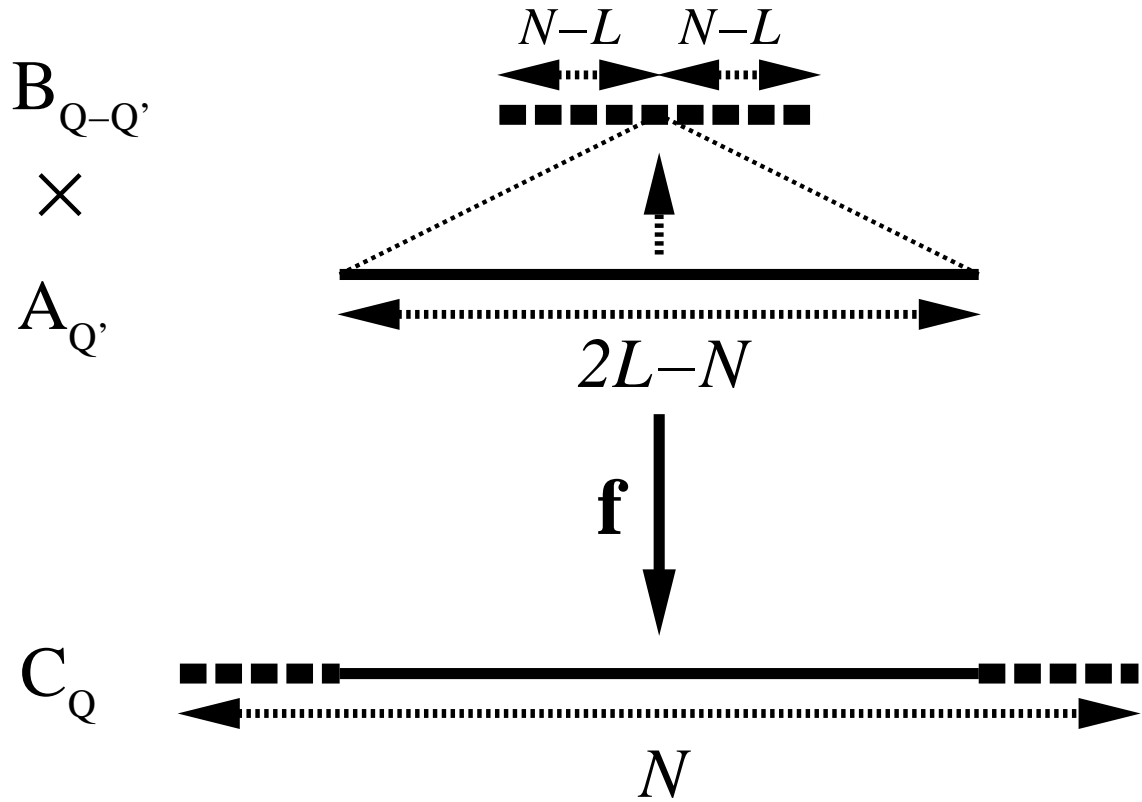}
}
\caption{Schematic illustration of the mapping $f$. A pair of sequences
from $B_{Q'}$ and $A_{Q-Q'}$ are combined to form a sequence from $C_Q$.}
\label{mapping}
\end{figure}

We first consider the following sets of random sequences, for
$N/2<L<N$ and arbitrary $Q$:
\begin{eqnarray*}
A_Q &=& \{\omega\in\Omega_{2L-N} : S_{2L-N}(\omega)=Q\}, \\
B_Q &=& \{\omega\in\Omega_{2(N-L)} : \mbox{Largest $Q$--segment
in $\omega$ has size $N-L$}\}, \\
C_Q &=& \{\omega\in\Omega_{N} : \mbox{Largest $Q$--segment
in $\omega$ has size $L$}\}.
\end{eqnarray*}
$A_Q$ is the set of all ($2L-N$)--step sequences with {\it total}
displacement (charge) $Q$. This set has $2^{2L-N}W_{2L-N}(Q)$
elements, where the function $W$ has been defined in
Eq.~(\ref{allcharge}). The set $B_Q$ contains all
($2N-2L$)--step sequences whose largest $Q$--segments are exactly
half as long as the whole sequence. By definition, there are
$2^{2(N-L)}P_{2(N-L)}(N-L,Q)$ such sequences.
Finally, $C_Q$ is our ``target set'' which consists of all
$N$--step sequences whose largest $Q$--segment has length $L$.
This set contains $2^N P_N(L,Q)$ sequences. We shall use the
sequences from the $A$-- and $B$--type sets to construct the
sequences of the ``target set'':
It is possible to construct a one-to-one onto mapping
\begin{equation}
 f:\bigcup_{Q'}\left(B_{Q'}\times A_{Q-Q'}\right) \mapsto C_Q,
\end{equation}
i.e., each sequence in $C_Q$ can be uniquely associated with a pair
of sequences from $B_{Q'}$ and $A_{Q-Q'}$ for some value of $Q'$,
and vice versa. The mapping $f$ is schematically shown in
Fig.~\ref{mapping}. Basically, the sequence from $A_{Q-Q'}$ is
inserted into the sequence from $B_{Q'}$ {\it at its midpoint}
to create a sequence in $C_Q$. After such an insertion we obtain a
sequence of length $2(N-L)+2L-N=N$, which contains a segment of
charge $Q-Q'+Q'=Q$ of length $(N-L)+(2L-N)=L$. Thus we created
an $N$--step sequence with a $Q$--segment of size $L$. From the
process of construction it is clear, that this is the {\it
largest} $Q$--segment in the sequence: if a larger $Q$--segment
had existed in the resulting chain, we could have reversed the
process by removing a segment of length $2L-N$ from the center
of the chain. This would have yielded a $2(N-L)$ step chain
whose largest $(Q-Q')$--segment was longer than half of its
entire length, contradicting the initial assumption regarding
the chain from the set $B_{Q-Q'}$.
The ``reversibility'' of the process also proves the
one--to--one correspondence between the sets. It should be
stressed, however, that this process requires that the midpoint
of the resulting $N$--step sequence is necessarily included in
the largest $Q$--segment. Thus, the proof is valid only for
$L\ge N/2$.

Since $A_{Q_1}$ and $A_{Q_2}$ are disjoint when
$Q_1\neq Q_2$, equating the number of elements in the domain and
range of $f$ gives the identity
\begin{equation}
\label{recrel}
P_N(L,Q)=\sum_{Q'}W_{2L-N}(Q-Q')P_{2(N-L)}(N-L,Q').
\end{equation}
Taking the continuum limit of the above equation, we replace the
probabilities $P$ by the probability density $p$, and the discrete
probability $W$ by its continuum (Gaussian) form which follows from
Eq.~(\ref{allcharge}) and obtain
\begin{equation}
\label{eintrel}
p(\ell,q)={1\over\sqrt{4\pi(2\ell-1)(1-\ell)}}\;
\int\limits_{-\infty}^{+\infty}\!\!dq'\,
{\rm e}^{-{\left(q-q'\sqrt{2(1-\ell)}\right)^2\over 2(2\ell-1)}}
p\left({1\over2},q'\right),
\end{equation}
where $q'\equiv Q'/\sqrt{N}$.
Since the equation is linear in the
function $p$, it cannot be used to determine proportionality
constants. (Since the equation is valid only for
$\ell\ge{1\over2}$, the normalization condition of $p$ cannot be used
either.) Eq.~(\ref{eintrel}) expresses an unknown function in an
interval of $\ell$s via the values of the same unknown
function at a particular point $\ell={1\over2}$. Despite these
limitations, Eq.~(\ref{eintrel}) can be utilized to
explain some properties of $p(\ell,q)$ and to extract
information using alternative methods, as will be explained
below. Before proceeding, we
note, that in the $\ell\to{1\over2}$ limit the Gaussian term
in the integrand of Eq.~(\ref{eintrel})
(the exponential term with the prefactor
$1/\sqrt{2\pi(2\ell-1)}$) becomes $\delta(q-q')$, and the
integral relation reduces to identity.

By integrating both sides of Eq.(\ref{eintrel}) over $q$,
we find a relation between the areas $A_\ell$, for $\ell>{1\over 2}$:
\begin{equation}\label{eqAl}
A_\ell\equiv\int\limits_{-\infty}^{+\infty}\!\!dq\,p\left(\ell,q\right)
={1\over\sqrt{2(1-\ell)}}\;
\int\limits_{-\infty}^{+\infty}\!\!dq\,p\left({1\over2},q\right)\ ,
\end{equation}
which confirms the observation from the MC data that for
$\ell>{1\over2}$, $A_\ell$ is proportional to $1/\sqrt{1-\ell}$.
The relation (\ref{eqAl}) provides a method for measuring the
otherwise unknown proportionality constant by detailed
calculation of probability density at $\ell={1\over2}$, i.e.
measurement of $A_{1/2}$.

In the $\ell\to1$ limit, the variable
$q'$ disappears from the exponent in Eq.(\ref{eintrel}),
and the relation reduces to
\begin{equation}\label{plimit}
p(\ell\to1,q)={A_{1/2}\over\sqrt{4\pi(1-\ell)}}\;{\rm e}^{-q^2/2}\ .
\end{equation}
This relation both confirms our contention that $p(\ell,0)$
has a square-root divergence $A/\sqrt{\pi(1-\ell)}$ with
$A\equiv{1\over 2}A_{1/2}$, and demonstrates that the
fixed--$\ell$ sections of the surface in Fig.\ \ref{plq}
approach a pure Gaussian shape when $\ell\to1$.

\begin{figure}
\centerline{
\epsfxsize=5.9truein
\epsffile{\dir/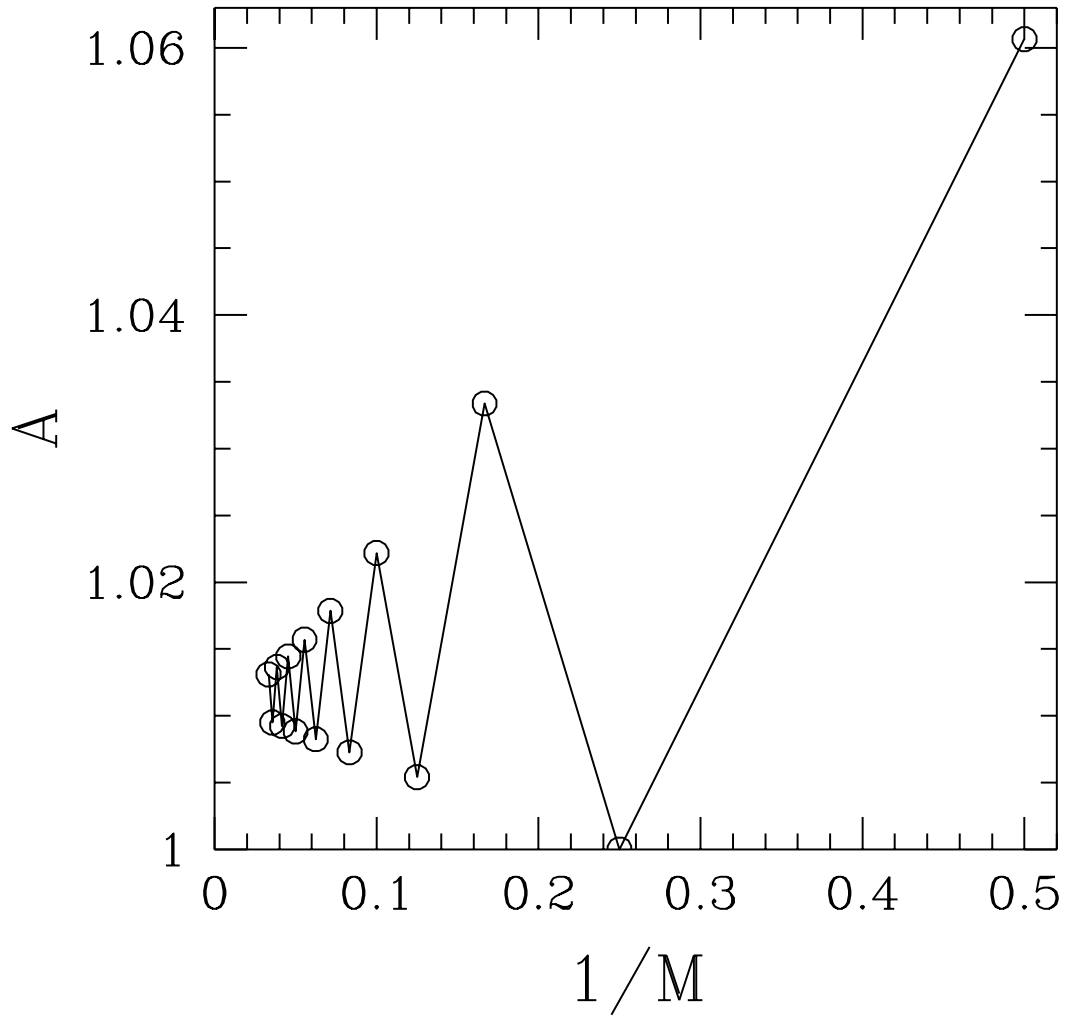}
}
\caption{Exact enumeration results for the determination of
the coefficient $A$.
The series $A(M)={1\over 2}\sum_Q P_M(M/2,Q)$ converges
to $A$ as $1/M\to 0 $.
}
\label{am}
\end{figure}

The proportionality coefficient of the square--root
divergence $A$ is simply related sum over $Q$ of the
probabilities for the largest $Q$--segment to be exactly half of
the length of the RS. By complete enumeration we calculated the
probabilities $P_M(M/2,Q)$ for all $Q$ and $M\le30$, and formed
the sums $A(M)\equiv{1\over 2}\sum_Q P_M(M/2,Q)$. (Only even
sequence lengths $M$ were used.)
The sums $A(M)$ converge to $A$ in the $M\to\infty$ limit.
Fig.~\ref{am} depicts the sequence of the estimates $A(M)$
plotted versus $1/M$. The extrapolation to $1/M=0$ provides
an estimate $A={1\over2}A_{1/2}=1.011\pm0.001$. This result is
consistent with the MC estimates of $A$, and has smaller error
bars. It is interesting to note, that despite the fact that
$A$ is almost unity, it is definitely larger than 1.

Finally, we note that the discrete relation in Eq.~(\ref{recrel})
can be used to produce exact analytical forms for $P_N$.
Consider cases when $L=N-M$ and $M$ is a small number.
Eq.~(\ref{recrel}) can be rewritten in the terms of $M$ as
follows:
\begin{equation}
P_N(N-M,Q)=\sum_{Q'}W_{N-2M}(Q-Q')P_{2M}(M,Q').
\end{equation}
Consider a case of, say, $M=2$. The function $P_4(2,Q')$ is
nonzero only for $Q'=0,\pm2,\pm4$, and can be easily found for
those cases by examining all random sequences of length 4.
The function $W_{N-4}(Q-Q')$ is known exactly for {\it
arbitrary} values of $N$ and $Q-Q'$. The sum over $Q'$ is
finite --- it contains only 5 terms, and therefore can be
performed. As a result we can find an exact expression for
$P_N(N-2,Q)$ for an arbitrary value of $N$. A Similar
procedure can be performed for $M=4$. Thus,
for {\it arbitrarily large} (even) $N$ we get
\begin{eqnarray*}
P_N(N-2,0)&=&2^{2-N}
  {(N-2)!\over\left[\left({N-2\over2}\right)!\right]^2}\ ,\\
P_N(N-4,0)&=&2^{1-N}
 {(N-8)!\over\left[\left({N-8\over2}\right)!\right]^2}
\cdot {91N^2-1186N+3576\over(N-4)(N-6)}\ .
\end{eqnarray*}
Unfortunately, the
expressions become increasingly complex with increasing $M$, and
it is not possible to use this method to determine the continuum
limit of $p(\ell,q)$.

We did not find analogous integral relations for
$\ell<{1\over2}$. Here, the situation is complicated by the
fact that, in a given sequence, there may be several longest
$Q$-segments that are disjoint.

\section{Extremal segments and the ``staircase problem''}
\label{related}

In this Section, we define a new problem in the theory of random
walks, related to two simultaneous walkers, and analyze it
detail. We derive the relation between this problem, and the
problem of extremal segments, and use this relation to investigate
the properties of $p(\ell,0)$ in the $\ell\to1$ limit.

\begin{figure}
\centerline{
\epsfxsize=5.9truein
\epsffile{\dir/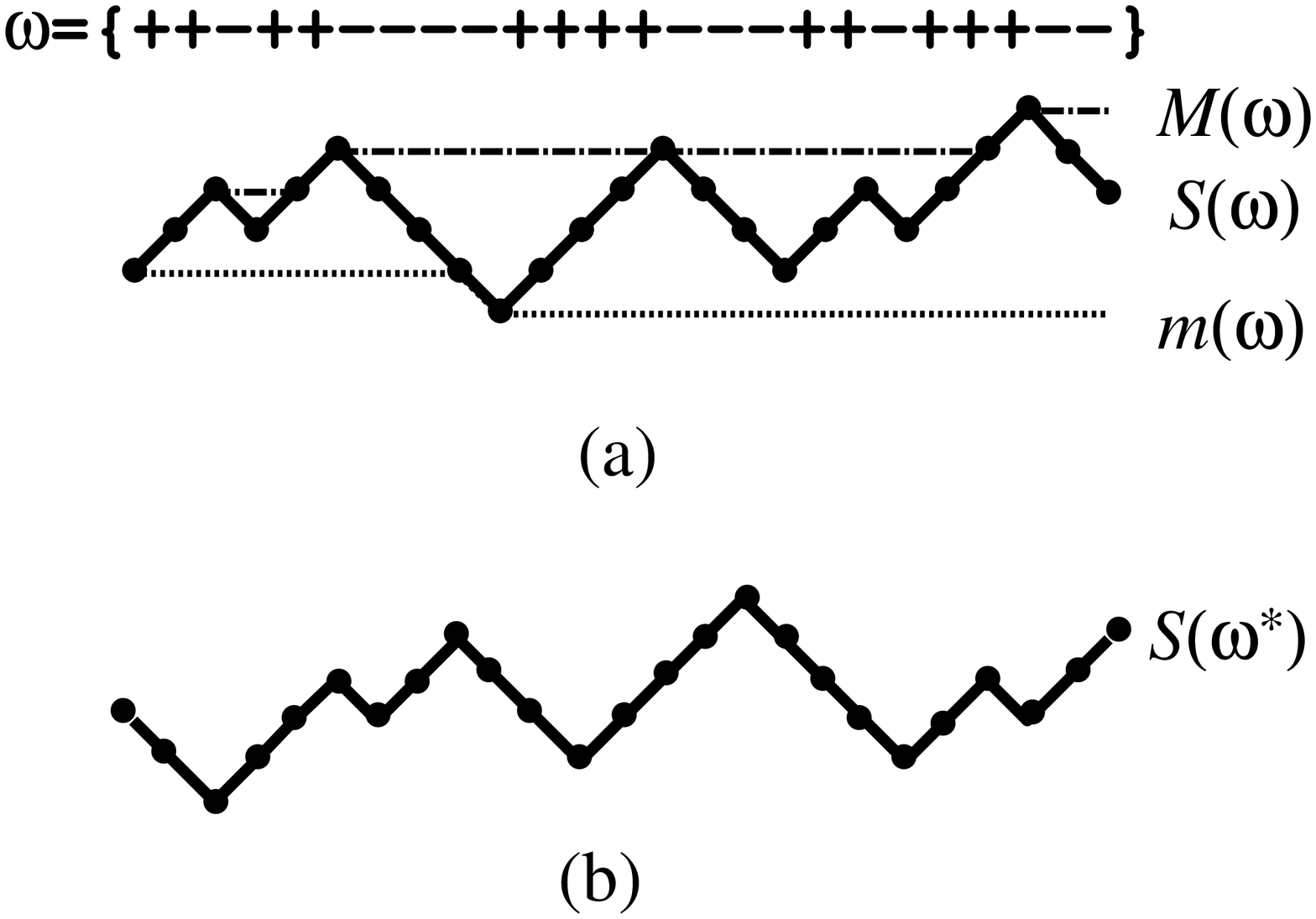}
}
\caption{(a) A sample RW $\omega$, depicted along with $S_i(\omega)$,
$M_i(\omega)$ and $m_i(\omega)$. (b) The conjugate sequence $\omega^*$.
}
\label{definitions}
\end{figure}

Consider a random sequence (walk) $\omega=\{q_1, q_2,\dots,q_N\}$. It
can be graphically represented by a plot of $S_i$ versus $i$,
where $S_i$ represents the total displacement of the $i$th step
from the origin of the walk. Let us define the following
variables:
\begin{eqnarray}
M_i(\omega)&\equiv&\max\left\{S_0(\omega),S_1(\omega),\cdots
S_i(\omega)\right\}, \\
m_i(\omega)&\equiv&\min\left\{S_0(\omega),S_1(\omega),\cdots
S_i(\omega)\right\}\ .
\end{eqnarray}
The variables $M_i$ and $m_i$ represent the maximal and minimal
coordinates achieved by the random walker up to (and including)
$i$th step. In Fig.~\ref{definitions}a, the dot--dashed and
dotted lines depict $M_i$ and $m_i$, respectively, corresponding
to a RS $\omega$ shown above the graph. (The corresponding
$S_i$ is depicted by the solid line.) The variable $M_i$
($m_i$) is a monotonic non--decreasing (non--increasing)
function of $i$ which graphically looks like an ascending
(descending) staircase.
One can also view $M_i$ and $m_i$ as two walls that contain
the entire RW. Initially the walls are
located at $M_0=m_0=0$, and they gradually separate from each
other: whenever the  random walker inside
reaches a wall and performs an additional step in the direction
of the wall, it pushes the wall to a new position thus increasing
the distance between the walls.

\begin{figure}
\centerline{
\epsfxsize=5.9truein
\epsffile{\dir/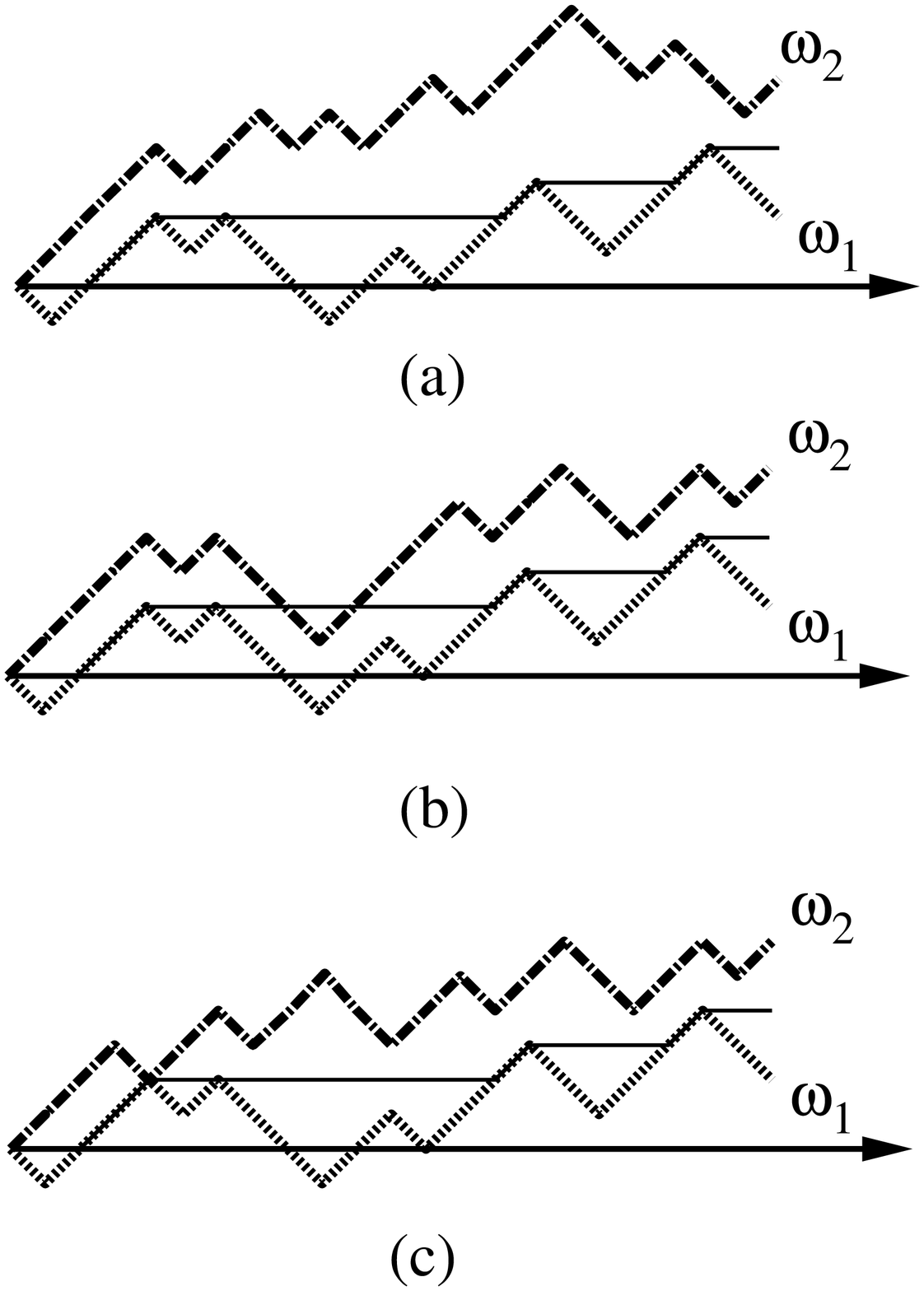}
}
\caption{Illustration of the probabilities $\phi_L$ and $\tilde\phi_L$.
Configuration (a) contributes to both, (b) to neither, and
(c) contributes to $\tilde\phi_L$ but not $\phi_L$.}
\label{3walks}
\end{figure}

Consider two RSs, $\omega_1$ and $\omega_2$,
selected from $\Omega_N$. We are interested in the probability
\begin{equation}
\phi_L={\rm Prob}(S_i(\omega_2)>M_i(\omega_1), 1 \leq i \leq L)
\end{equation}
that the path $\omega_2$ remains {\it above} the maximum point
of $\omega_1$ that far, for the first $L$ steps. The dotted line in
Fig.~\ref{3walks} depicts the RS $\omega_1$, which generates the
staircase (solid line) that the RS $\omega_2$ is supposed
to remain entirely above of. We denote the determination
of $\phi_L$ as the ``staircase
problem." The dot--dashed line in Fig.~\ref{3walks}a depicts a permitted
$\omega_2$, while dot--dashed lines in Figs.~\ref{3walks}b and
\ref{3walks}c show examples of forbidden cases.
(Analogously, one can define of problem of
RW staying below $m_i$, and a problem of RW staying either above
$M_i$ or below $m_i$, i.e. staying outside the walls
pushed by the RS $\omega_1$.)
Every step of the staircase begins when the RS $\omega_1$
arrives to that particular maximal value of $S$ for the first time.
The step ends when the sequence exceeds that value for the first
time. The sizes of the these steps are independent of each other,
and their distribution
is given by the first arrival time to index 1, i.e.,
Prob(Size of a step$=k)=k^{-1}{\rm Prob}(S_k=1)\sim k^{-3/2}$.
(For a general expression of first arrival times see
Ref.~\cite{rchandra}.) This probability is normalizeable, but
the {\it mean} step size is divergent.

\begin{figure}
\centerline{
\epsfxsize=5.9truein
\epsffile{\dir/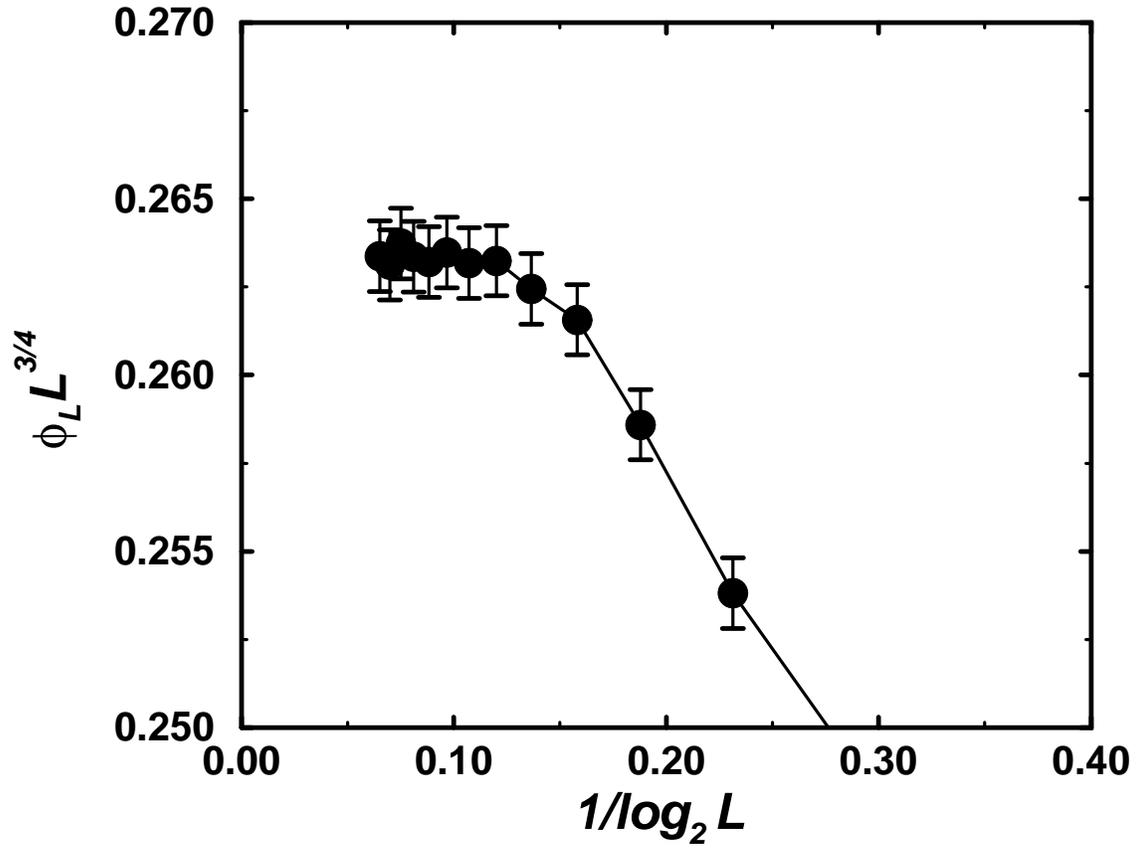}
}
\caption{Numerical demonstration of the power law relation
$\phi_L\protect\sim L^{-3/4}$, and the determination of
the constant $C_\phi$.}
\label{Cphi}
\end{figure}

The probability $\phi_L$ of $\omega_2$
staying above the staircase after $L$ steps decreases with
increasing $L$.
It is easy to put loose upper and lower bounds to $\phi_L$:
(i) $\omega_2$ needs to remain above the origin up to the
$L$th step, since $M_i(\omega_1)\geq 0$. Thus, $\phi_L$
decays faster than $L^{-1/2}$, which is the asymptotic
behavior of the probability of never
returning to the origin given by Eq.~(\ref{identity}).
(ii) The condition is satisfied if $\omega_1$ remains
completely below the origin and $\omega_2$ remains
above the origin up to step $L$. Therefore, $\phi_L$ decays
slower than $L^{-1}$. Given these bounds, it is reasonable
to expect an asymptotic power law for $\phi_L$:
\begin{equation}
\lim_{L\to\infty}\phi_L=C_\phi L^{\alpha-1},
\end{equation}
where $0\leq\alpha\leq {1\over2}$. We will later argue that
$\alpha={1\over4}$. We performed a MC investigation of the
staircase problem for $L$ ranging from 10 to 40960 and sample
sizes of about $3\times10^5~L^{3/4}$
(yielding approximately $3\times10^5$ survival events for
each $L$), and confirmed this particular value of
$\alpha$ to within one percent.
Fig.~\ref{Cphi} shows $\phi_L\times L^{3/4}$ as a function of
$1/\log_2(L)$.
The fact that this combination remains independent of
$L$ when $L\to\infty$ demonstrates the assumed power law.
The points on the graph
provide successive estimates of the prefactor $C_\phi$; the
error bars indicate  statistical uncertainties
(one standard deviation) for each $L$.
We estimate the asymptotic value of the coefficient as
$C_\phi=0.263\pm0.001$.

\begin{figure}
\centerline{
\epsfxsize=5.9truein
\epsffile{\dir/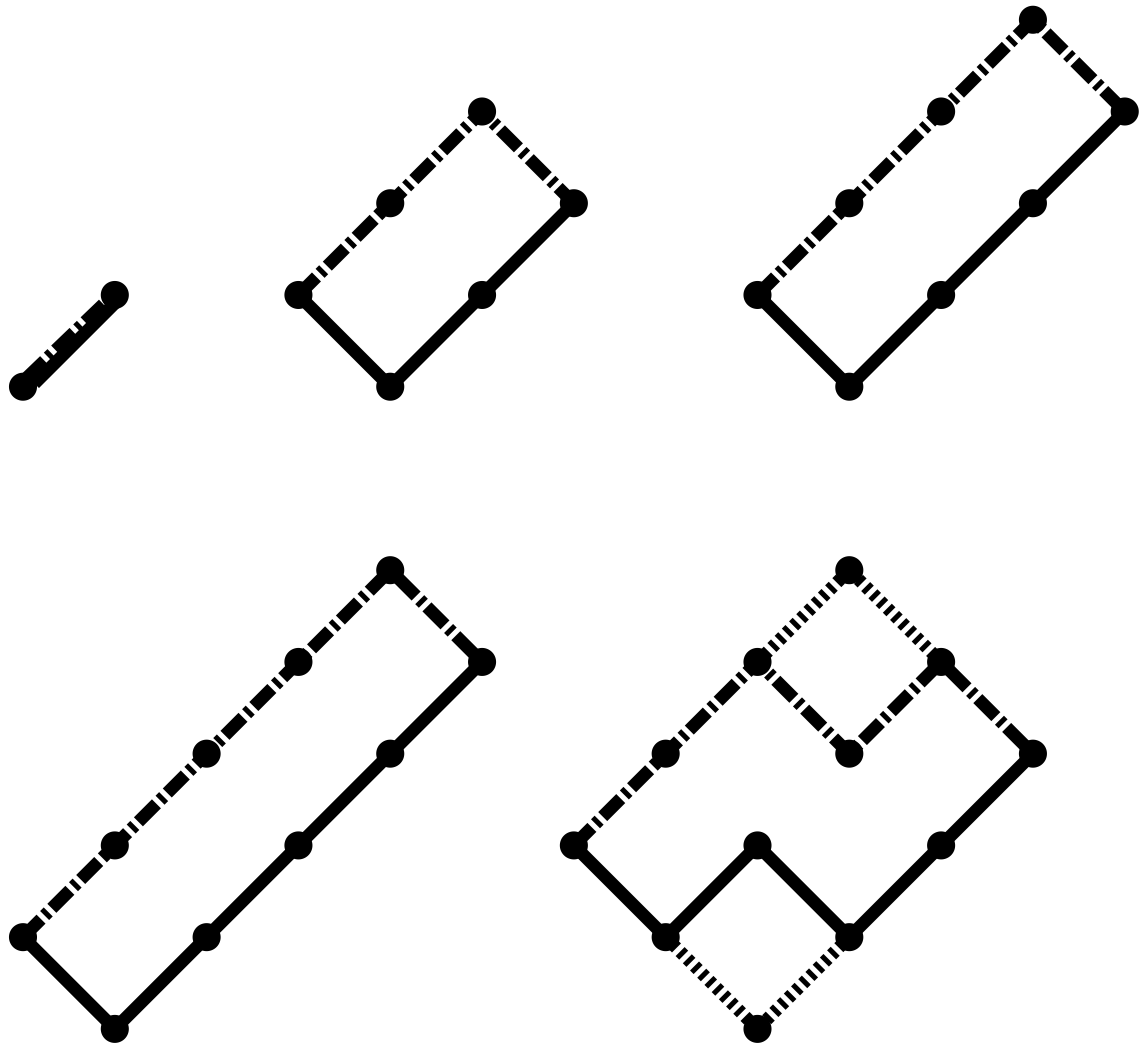}
}
\caption{The first few configurations used to generate the
statistical weights $f_L$ (see text).}
\label{fterms}
\end{figure}

\begin{figure}
\centerline{
\epsfxsize=5.9truein
\epsffile{\dir/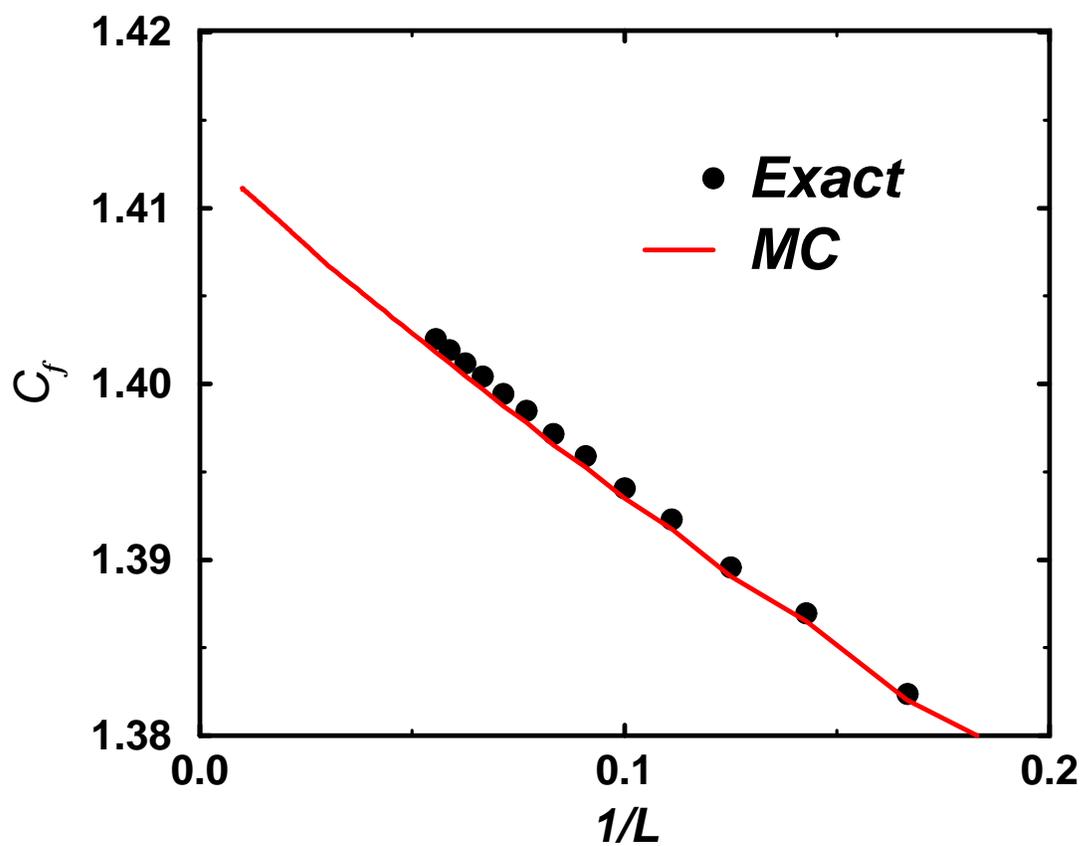}
}
\caption{The value for
$C_f=\lim_{L\to\infty}(1-\sum_i^L f_i)^{-1}$ is calculated
by keeping a finite number of terms in the series and
extrapolating to $1/L=0$. Both exact and
Monte Carlo data are shown. The MC data is obtained by starting
with single ensemble of $10^8$ RWs, thus the data
points are not statistically independent.}
\label{Cf}
\end{figure}

A very closely related probability distribution is
\begin{equation}
\tilde\phi_L={\rm Prob}(S_i(\omega_2)>M_{i-1}(\omega_1), 1\leq i\leq L),
\end{equation}
i.e., this time the two paths are allowed to meet at positions where
$\omega_1$ has reached a new maximum. Figs.~\ref{3walks}a and
\ref{3walks}c both correspond to the permitted events in the
definition of $\tilde\phi_L$.
Now let
\begin{equation}
f_L={\rm Prob}\left(S_i(\omega_2)>M_i(\omega_1),1 \leq i \leq L-1;\;
S_L(\omega_2)=S_L(\omega_1)\right)
\end{equation}
denote the probability of such a meeting occurring for the
first time at step L.
Meeting at the $L$th step represents an extremely simple event, i.e.,
despite the fact that we are considering the behavior of {\it two}
random walkers, it is easy to construct all possible cases for short $L$.
In Fig.~\ref{fterms}, the solid and dot--dashed lines represent
$\omega_1$ and $\omega_2$, respectively, for $L=$1, 3, 4 and 5. We can
see that there is a single possibility for $L=1,3,4$ and
five possibilities for $L=5$. (The diagram in the bottom right represents
4 different cases; the dashed lines indicate the alternative segments
in both $\omega_1$ and $\omega_2$.) $f_L$ is simply equal
to $2^{-L}$ (probability of a single diagram) multiplied by
the number of distinct such diagrams. Since $\{f_i\}$
is a rapidly converging series, we can easily evaluate the
infinite sum $\sum_i f_i$ to a high accuracy by summing
the first few terms. (The convergence
of the infinite series $\sum_i f_i$ can be easily seen from
the fact that it is bounded from above by the probability that
$\omega_1$ is at a maximum when the two RWs meet for the
first time.)
We can use the probabilities $f_i$ to relate $\tilde\phi_L$
to $\phi_L$ via the following relation:
\begin{equation}\label{phitophi}
\tilde\phi_L=\phi_L+\sum_{L_1=1}^L f_{L_1}\phi_{L-L_1}
+\sum_{L_1=1}^L\sum_{L_2=1}^{L-L_1} f_{L_1}f_{L_2}\phi_{L-L_1-L_2}+\cdots.
\end{equation}
Fast decay of $f_L$ with increasing $L$, allows the replacement of
$\phi_{L-L_1}$, $\phi_{L-L_1-L_2},\cdots$ in  Eq.(\ref{phitophi})
by $\phi_L$ in the $L\to\infty$ limit, leading to
\begin{equation}
\tilde\phi_L\mathop{=}_{L\to\infty}\frac{1}{1-\sum_i f_i} \phi_L
\equiv C_f\phi_L.
\end{equation}
The coefficient $C_f$ can be calculated to high accuracy by
summing up the series $\{f_i\}$. We have obtained the value
$C_f=1.413\pm0.005$ by extrapolating from finite sums of
$f_i$, which we have obtained exactly for $L$ up to 18,
and up to $L=100$ using a Monte Carlo method.
The results are shown in Fig.~\ref{Cf}.

\begin{figure}
\centerline{
\epsfxsize=5.9truein
\epsffile{\dir/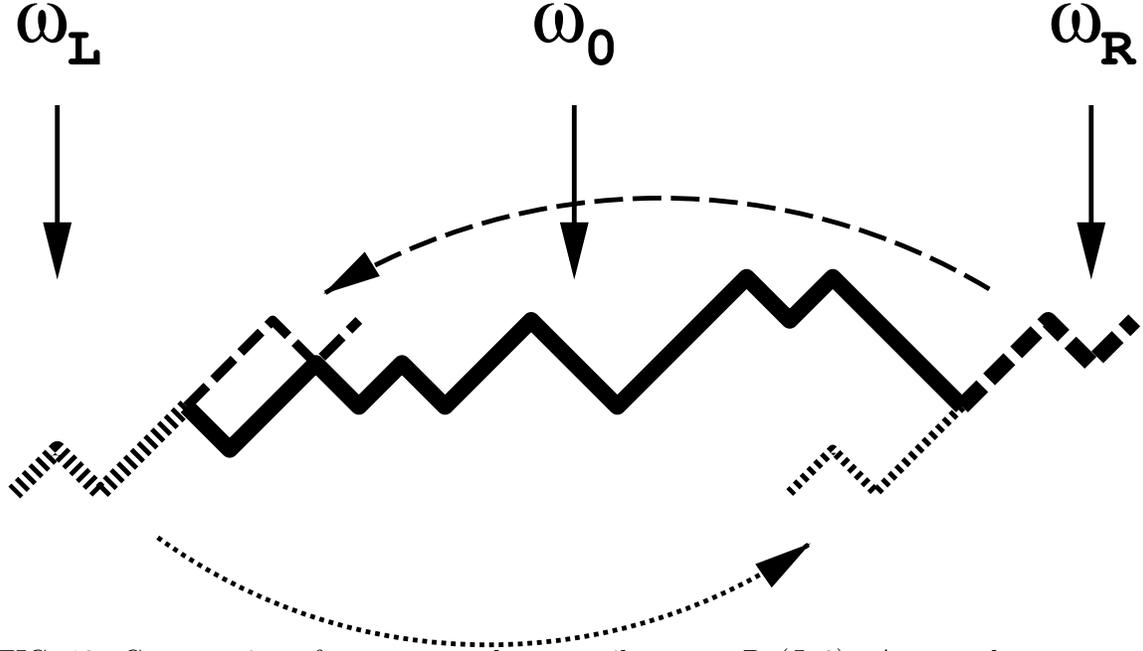}
}
\caption{Construction of a sequence that contributes to $P_N(L,0)$.
A neutral segment $\omega_0$ of length $L$ is augmented by two segments
$\omega_L$ and $\omega_R$, such that $\omega_R$ ($\omega_L^*$) always
stays above the maximum of $\omega_0$ ($\omega_0^*$). $\omega_R$
is allowed to touch a new maximum of$\omega_0$, since this only
produces neutral segments of length $L$ which are to the right of
$\omega_0$.
}
\label{twowalkers}
\end{figure}

Finally, we are in a position to discuss the connection of the
staircase problem to the problem of our main interest.
For simplicity, let us only consider $P_N(L,0)$ in the $L/N \to 1$
limit and examine all RSs with $S_N(\omega)>0$ whose largest
neutral segments are $L$ steps long. To construct such a sequence,
we can start with a neutral segment $\omega_0$ of size $L$,
depicted by a solid line in Fig.~\ref{twowalkers}. This segment is completed
into the $N$--step RS by adding pieces to its two ends (thick
dashed and dotted lines in Fig.~\ref{twowalkers}), in such a way {\it that a
larger neutral segment is not created}. In order to avoid
overcounting when there is more than one
largest neutral segment, we can for example require that the initially
selected segment is the {\it leftmost} of all largest segments.
Let $L'$ be the size of the piece $\omega_R$ added to the RHS
of $\omega_0$. (The LHS piece $\omega_L$ will then have length $N-L-L'$.)
To avoid creating a larger neutral segment which begins somewhere
inside $\omega_0$ and ends somewhere inside $\omega_R$, the
sequence $\omega_R$ must remain
above the staircase generated by the successive maxima of
$\omega_0$, i.e. if the sequence $\omega_R$ is translated to the
beginning of the sequence $\omega_0$ (as depicted by the thin dashed
line in Fig.~\ref{twowalkers}) they must satisfy
conditions defined in the staircase problem. Similar
restrictions apply to the segment $\omega_L$; however, this time
both $\omega_0$ and $\omega_L$ should be viewed ``backwards''
(thin dotted line in Fig.~\ref{twowalkers}).
(Formally, for any sequence $\omega$ it is convenient to define
a {\it conjugate sequence} $\omega^*$, which consists of the
elements $q_i$ of $\omega$ written in reverse order, as
illustrated in the Fig.~\ref{definitions}b.
The conjugate of
a given path can be obtained by rotating the original path by
$180^\circ$ around the axis normal to the plane.
Thus, the
staircase conditions have to be satisfied between the sequences
$\omega_0^*$ and $\omega_L^*$.) Since $\omega_0$ is
a neutral segment, its elements are not completely
independent, while our original definition of the staircase
problem required the presence of two completely random
sequences. However, when $N-L\ll N$, the two ends
of $\omega_0$ can be treated approximately as independent RSs,
and they become completely independent in the $(N-L)/N\to0$
(i.e., $\ell\to1$) limit. Finally, we notice that the above
requirements were somewhat over--restrictive: we {\it are allowed} to
create neutral segments exactly of length $L$ between
$\omega_0$ and $\omega_R$, and therefore the probability will be
described by $\tilde\phi_L$ rather than by $\phi_L$. The
segment $\omega_L$, however, must satisfy probabilities described
by $\phi_L$ because we initially required that the neutral
segment created by $\omega_0$ is the {\it leftmost} segment
in the RS. This yields
\begin{equation}
P_N(L,0)=2\sum_{L'=0}^{N-L}\phi_{N-L-L'}W_{L}(0)\tilde\phi_{L'},
\end{equation}
the factor 2 coming from RWs with $S_N(\omega)<0$.
Finally, taking the sum
over $L'$ in the large $N$ limit, we obtain (for even L)
\begin{eqnarray}
\lim_{L/N\to 1^-}P_N(L,0)&=&
\frac{2C_\phi^2C_f}{(N-L)^{1/2-2\alpha}}
\sqrt\frac{2}{\pi L(N-L)}
\int_0^1 \frac{d\ell'}{[\ell'(1-\ell')]^{1-\alpha}} \nonumber \\
&=& \frac{\Gamma(\alpha)\Gamma(\alpha)}{\Gamma(2\alpha)}
\frac{2C_\phi^2C_f}{(N-L)^{1/2-2\alpha}}
\sqrt\frac{2}{\pi L(N-L)}.
\end{eqnarray}
In the above, $\Gamma(x)$ is the gamma (factorial) function.
This result has several remarkable consequences:
First of all, this result suggests that $p(\ell,0)$ has a
well-behaved continuum limit only if $\alpha=1/4$.
This implies that $\phi_L\sim L^{-3/4}$,
a result we have not  yet found in the literature.
Knowledge of $C_\phi$ and $C_f$ now enables an
independent calculation of the proportionality coefficient
$A$ through the relation
$A=\sqrt{2}C_\phi^2C_f[\Gamma(1/4)]^2
/\Gamma(1/2)=1.025\pm0.015$.
Although it is slightly larger and less accurate,
this result is consistent with other estimates of $A$.

\section{Higher Dimensions}\label{highd}

The fact that $p(\ell,0)$ has a singularity at $\ell=1$ is
a consequence of the fact that a RW in one dimension
returns to its starting position very often. Thus, it is
clear that the behavior of $p(\ell,0)$ depends strongly
on the dimensionality of the RW. In order to investigate
this, we have generalized the original problem to RWs
on a $d$--dimensional hypercubic lattice. Now the ``elementary
charge'' (scalar) of the one dimensional problem is replaced
by an elementary step (vector) between neighboring sites
on that lattice along one of $2d$ possible directions, and
there are $(2d)^{N}$ possible $N$--step walks. (We cannot
use the analogy with the sequence of charges, anymore.)
The probability distribution $P_N(L,Q)$ can be
easily generalized:
\begin{eqnarray*}
P_N(L,Q) &\to& P_N^{(d)}(L,{\bf Q}), \\
p(\ell,q) &\to& p^{(d)}(\ell,{\bf q}),  \\
A_\ell &\to& A^{(d)}_\ell,
\end{eqnarray*}
where ${\bf Q}=(Q_1,\cdots,Q_d)$ is now the $d$-dimensional
displacement of a segment in the RW, and
${\bf q}=(dN)^{-1/2}{\bf Q}$.

Most of the arguments used to explore the features of
one--dimensional RWs can be applied with minor changes to
the $d$--dimensional walks. As an example, let us consider
the qualitative derivation of the asymptotic properties
of $p(\ell,{\bf 0})$ in the $\ell\to1$ limit as derived
for the $d=1$ case in Sec.~\ref{secapprox}: As in the
one--dimensional case we may assume that the length of the
longest loop can be approximately thought of as a function of
the {\it overall displacement} ${\bf Q}_o$ (end--to--end vector)
of the entire walk. Under such assumption we expect
$L\approx N-a|{\bf Q}|^2$, which is analogous to the
one--dimensional case, except for the overall charge $Q_o$ that is
replaced by the {\it modulus} (length) of the vector
${\bf Q}_o$. The generalization of Eq.(\ref{eqapproxp}) to
$d$--dimensions is
\begin{equation}\label{eqapproxpd}
p^{(d)}(\ell,{\bf 0})\approx{N\over 2}{\cal W}^{(d)}_N
(|{\bf Q}_o(L)|)
\left|{d|{\bf Q}_o|\over dL}\right|\ ,
\end{equation}
where the one--dimensional $W_N(Q_o)$ of Eq.~(\ref{eqapproxp}) has been
replaced by ${\cal W}^{(d)}_N(|{\bf Q}_o|)$,
which is the probability
that the {\it length} of a $d$--dimensional end--to--end vector
of an $N$--step RW is $|{\bf Q}_o|$. Near ${\bf Q}_o={\bf 0}$
this probability is proportional to $N^{-d/2}|{\bf Q}_o|^{d-1}$.
Substituting, this expression to Eq.(\ref{eqapproxpd}) and
using the relation between $L$ and $|{\bf Q}_o|$ we find
$p(\ell,{\bf 0})\sim (1-\ell)^{(d-2)/2}$. Thus, we expect
the probability density to approach a constant in the $\ell\to1$
limit in $d=2$, and to decay to zero as $\sqrt{1-\ell}$ in $d=3$.

\begin{figure}
\centerline{
\epsfxsize=2.9truein
\epsffile{\dir/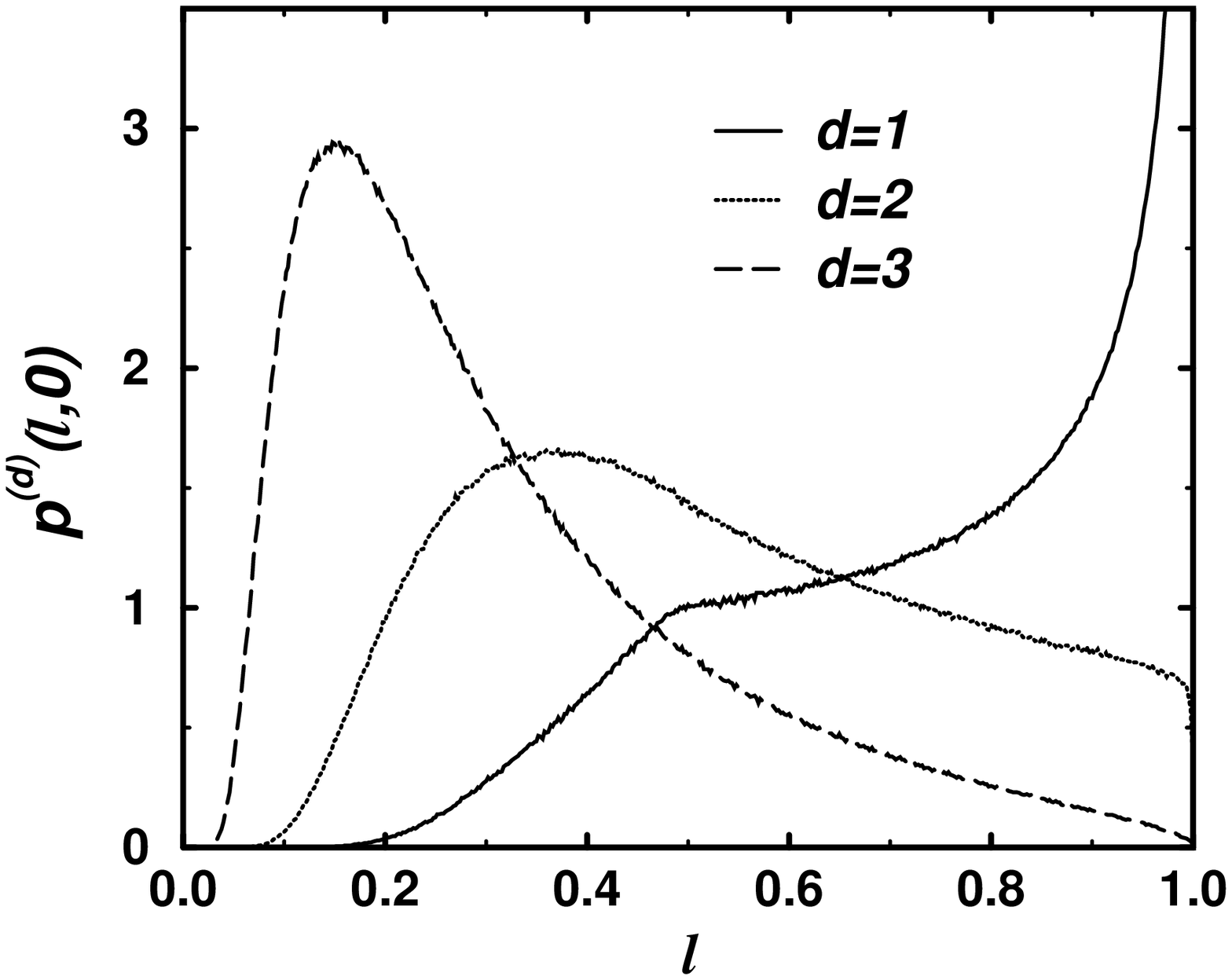}
\qquad
\epsfxsize=2.9truein
\epsffile{\dir/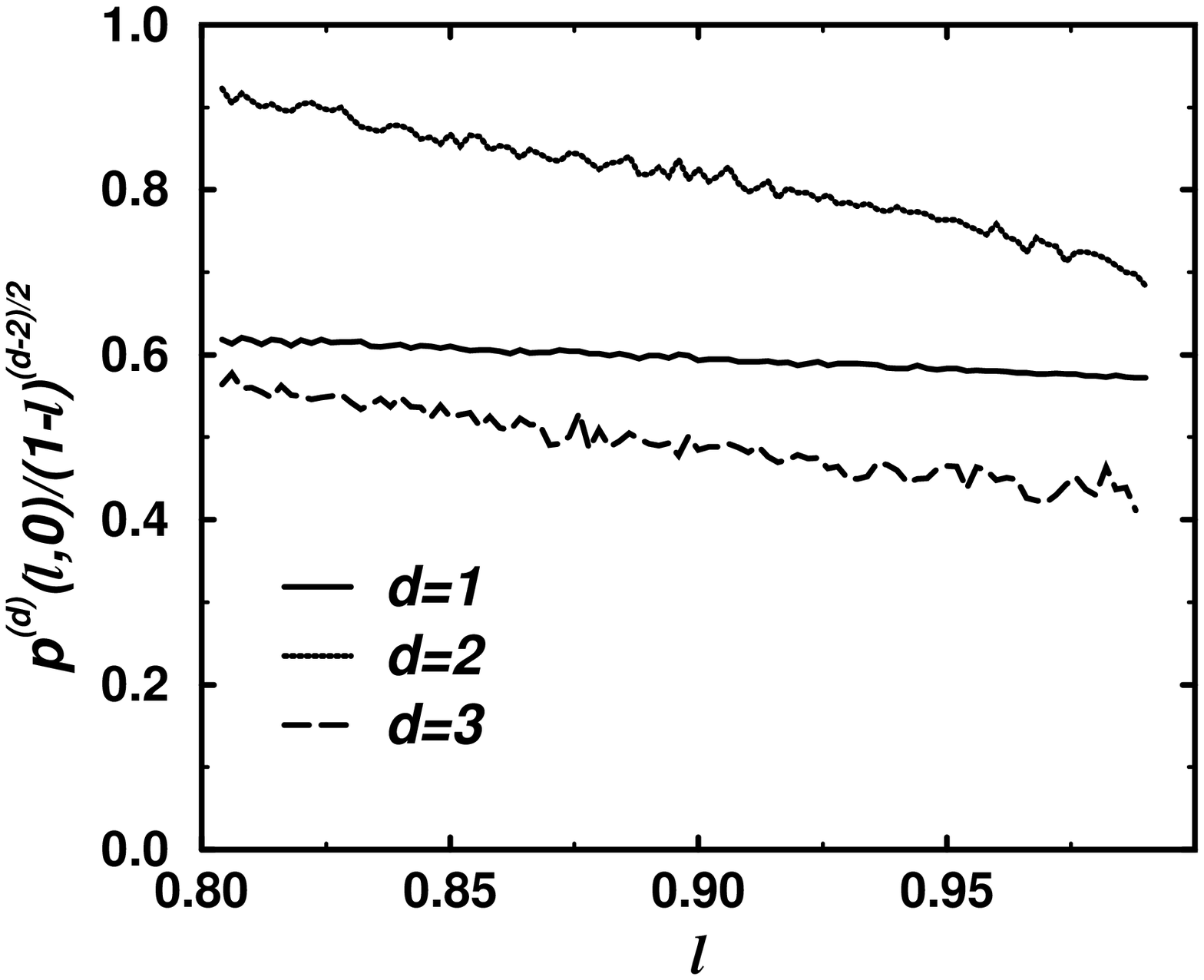}
}
\caption{{\it Left:} The distribution functions
$p^{(d)}(\ell,{\bf 0})$ in
1, 2 and 3 dimensions. {\it Right:} The $\ell\to 1$ limit of the
distributions.}
\label{alldimens}
\end{figure}

The relations which have been demonstrated from an approximate
argument above can be proven exactly by generalizing
Eq.(\ref{eintrel}) to $d$ dimensions. The generalization
is straightforward and leads to the form
\begin{equation}
p^{(d)}(\ell,{\bf q})={1\over 2(1-\ell)}
\left({1-\ell\over\pi(2\ell-1)}\right)^{d/2}\;
\int\limits_{-\infty}^{+\infty}\!\!d^dq'\,
{\rm e}^{-{\left|{\bf q}-{\bf q'}
\sqrt{2(1-\ell)}\right|^2\over 2(2\ell-1)}}
p\left({1\over2},{\bf q'}\right).
\end{equation}
Thus, for the $\ell\to 1$ limit we obtain
\begin{equation}
p^{(d)}(\ell\to 1,q)={A^{(d)}_{1/2}\over
2\pi^{d/2}}(1-\ell)^{d-2\over 2}e^{-|{\bf q}|^2/2}.
\end{equation}
Fig.~\ref{alldimens} depicts $p^{(d)}(\ell,0)$
for $d=1,2$ and 3, obtained from MC simulations $N=1000$ with
samples of $10^8$, $10^6$ and $10^6$ RWs, respectively.
The peak of the distribution shifts
towards $\ell=0$ as the dimensionality is increased.
Fig.~\ref{alldimens} also demonstrates the verification
of the form $p^{(d)}(\ell\to 1,0)\sim(1-\ell)^{(d-2)/2}$
for these dimensions.

The asymptotic relation described above
assumes that $A^{(d)}_{1/2}$ does not vanish.
Note that
\[
A^{(d)}_{1/2}=\lim_{N\to\infty} {N^{1-d/2}\over 2}
\sum_{\bf Q}P^{(d)}_N(N/2,{\bf Q}).
\]
For each sequence $\omega$, there are at most $N$
nonzero terms in the summation over {\bf Q}, and
$P^{(d)}_N(L,{\bf Q})<1$ since it is a probability.
Thus, $A^{(d)}_{1/2}\leq \lim_{N\to\infty}N^{2-d/2}$.
This implies that in dimensions higher than 4,
$p^{(d)}(\ell,{\bf q})=0$ for $l>1/2$.
It is easy to understand why $d=4$ is a special dimension in
this problem: It is known from the study of the self--avoiding
random walks\cite{degen} that large loops are absent in space
dimensions $d>4$. Thus we expect that in terms of the reduced
variable $\ell$, all loops will have reduced ``length''
$\ell=0$, i.e.,  $p^{(d)}(l,{\bf 0})=\delta(\ell)$ in this
regime.

\begin{figure}
\centerline{
\epsfxsize=2.9truein
\epsffile{\dir/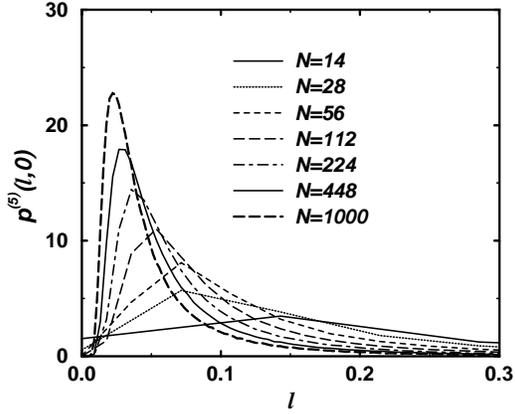}
\qquad
\epsfxsize=2.9truein
\epsffile{\dir/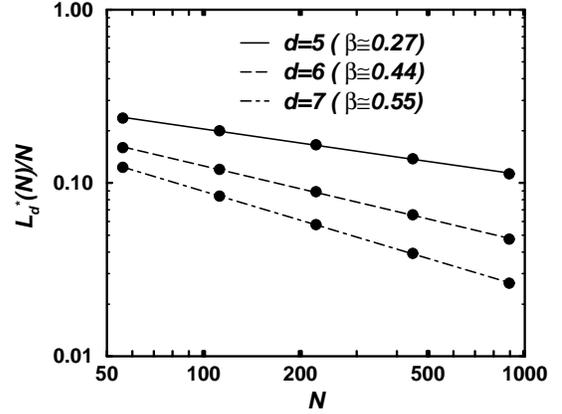}
}
\caption{{\it Left:} The distribution function
$p^{(5)}(\ell,{\bf 0})$ approaches a delta function with
increasing $N$. {\it Right:} The 90\% threshold
$L_d^*(N)$ scales with the RW size $N$, the slope in the
log-log plot gives $\beta_d$ for $d=5,6,7$.}
\label{dimhigh}
\end{figure}

While we expect an asymptotic probability density $\delta(\ell)$
for $d>4$, it should be noted that for {\it finite} $N$ the
probability $P_N(L,{\bf 0})$ is a monotonically increasing
function of $L$ for small values of $L$. Therefore, the
probability density $p(\ell,{\bf 0})$ measured for finite $N$
will have a peak at finite (small) value of $\ell$. As $N$
increases the entire distribution should drift towards
$\ell=0$. Fig.~\ref{dimhigh} depicts such a trend for $d=5$.
A convenient measure of such behavior is calculation
of value of $L$ such that most of the statistical weight
corresponds to loops shorter than the threshold value.
We verified the approach of the distribution to a
$\delta$--function by examining
the finite size scaling of the 90\% threshold
$L_d^*(N)$, defined through $\sum_{L=0}^{L_d^*(N)}
P^{(d)}_N(L,{\bf 0})=0.9$. This means that
10\% of the time, there is a loop larger than $L_d^*(N)$
in a $d$--dimensional RW of size $N$.
We examined the cases $d=5$, 6, and 7 using the MC method
described in the Appendix, for
$N$ ranging from 14 to 896 and sample sizes of $10^6$.
The threshold lengths are also shown in Fig.~\ref{dimhigh}.
We find that
$L_d^*(N)\sim N^{1-\beta_d}$, where
$\beta_5\approx0.27, \beta_6\approx0.44$ and
$\beta_7\approx0.55$. Since the exponents $\beta_d$
are positive, the threshold in the terms of the reduced variable
$\ell_d^*(N)\sim N^{-\beta_d}$ vanishes with increasing $N$.

\begin{figure}
\centerline{
\epsfxsize=2.9truein
\epsffile{\dir/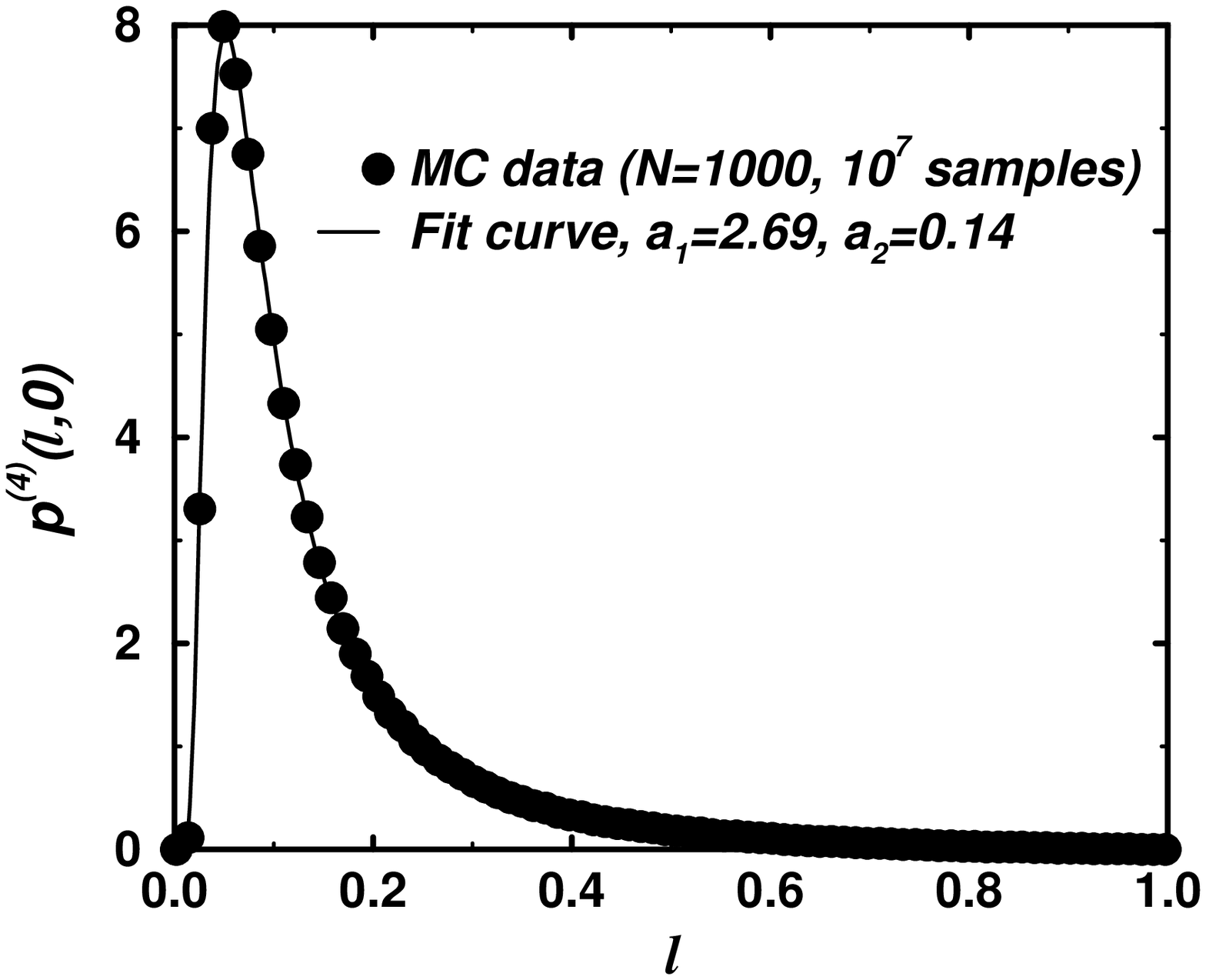}
\qquad
\epsfxsize=2.9truein
\epsffile{\dir/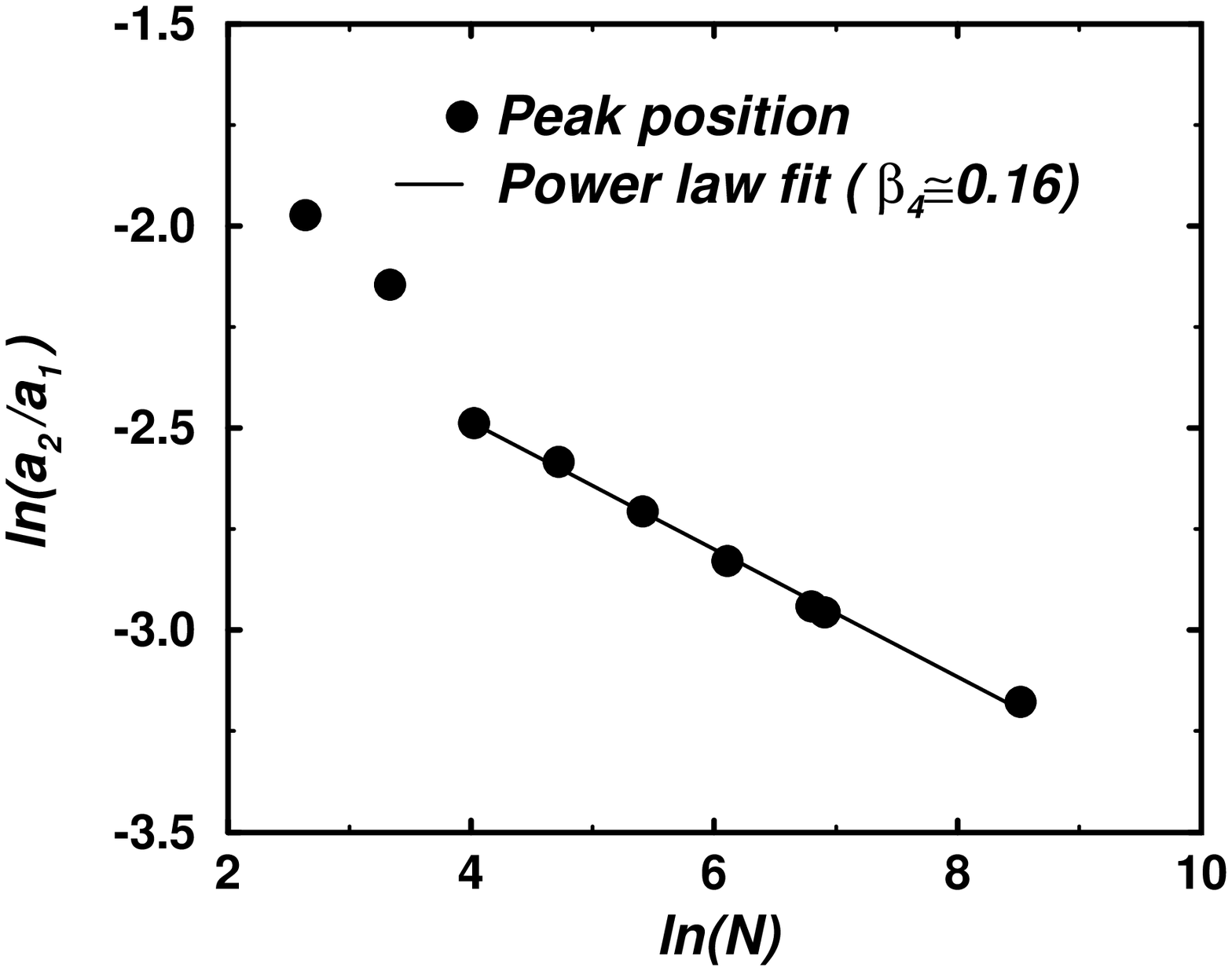}
}
\caption{{\it Left:} The distribution function
$p^{(4)}(\ell,{\bf 0})$ is fitted very well with a
function of the form $\ell^{-a_1}\exp(-a_2/\ell)$.
{\it Right:} $\beta_4$ is determined from the finite size
scaling of the peak positions as approximately 0.16.}
\label{dim4}
\end{figure}

The above arguments do not provide a definite answer for
the borderline dimension of $d=4$.
(The reader is reminded that the self--avoiding walk problem
at the critical dimension $d=4$ slightly differs
from the $d>4$ cases: e.g., regular power laws
are modified by logarithmic corrections.)
Through MC calculations with up to $10^7$ RW samples and
values of $N$ up to 5000, we find that in $d=4$ the
{\it entire} distribution
$p^{(d)}(\ell,{\bf 0})$ can be fitted very well to the
form $\ell^{-a_1}e^{-a_2/\ell}$, for finite values of $N$.
Fig.~\ref{dim4} depicts such curve for $N=1000$. (The sample size
is $10^7$.)
The peak position $a_2/a_1$ approaches 0 either logarithmically
$(a_2/a_1\sim1/\ln N)$, or with a very small power of $N$, i.e.
$a_2/a_1\sim N^{-\beta_4}$ where
$\beta_4\approx 0.16$. Thus, the distribution still converges
to a delta function in the continuum limit.
Although the qualitative behavior of $p^{(d)}(\ell,{\bf 0})$ is
easily understood, it would be interesting to obtain a
quantitative understanding of the distribution, especially
at the borderline dimension of four.

\section{Discussion}

The problem of extremal segments originated from the desire to
consider a simplified description of the ground states of
randomly charged polymers. We used MC, exact enumeration and
analytical techniques to analyze the problem, and our results
provide convenient tools for a semi--quantitative analysis of the
the ground states of PAs. In particular, we show that a
``typical'' RS contains very large neutral segments, i.e. it is
possible to construct a ground state from a single very large
blob with relatively short ends of the chain dangling outside
the blob.

Besides the original motivation, the problem of extremal
segments is interesting in its own right. It looks like one of
the classical problems of random walks and,
nevertheless, is highly non--trivial, and the results indicate a
solution with very rich and unexpected structure. The problem
can be related to other interesting problems of the RWs, such as
the ``staircase problem.'' While several features of the problem
have been established analytically, we did not find a complete
analytical solution of the problem. We think that such a solution
is possible and  further attempts of finding it are worthwhile.
Generalization of the problem to arbitrary space dimension $d$
is not related to the original problem of charged polymers,
nevertheless interesting in its own right.

The numerical ``proof'' of the continuum limit in our work was
limited to a particular class of RWs, in which a unit
displacement appears at each step. Within that class we
presented evidence of a continuum limit where
the properly scaled functions become independent of $N$. It
would be interesting to perform a numerical test of the
``universality'' of the solution for a broader class of RWs.
It may be possible to prove the universality of the continuum
limit by attempting to perform a renormalization--group--like
treatment of the problem, i.e. attempting to define the problem
in the limit where the RW becomes a true Gaussian walk (walk of
idealized Brownian particle). This limit, however, is far from
being trivial. In particular the definition of what is called a
loop (i.e. how close two different points of the walk should be
located so that the segment will be called a closed loop) presents
a non--trivial problem in the continuum limit. Such
short distance scale can undergo a non--trivial scaling,
similarly to the excluded volume parameter in the treatment of
self--avoiding walks. A different approach to the question
of universality may begin from an expansion of the solution
near the dimension $d=4$, as in the treatment of
self--avoiding walks.

\acknowledgments

We would like to thank M.~Kardar and A.~Yu.~Grossberg
for helpful discussions.
This work was supported by the US--Israel BSF grant
No. 92--00026, and by the NSF through grant Nos.
DMR--94--00334 (at MIT's CMSE) and DMR--91--15491 (at Harvard).

\begin{appendix}

\section*{Numerical Methods}

In this appendix, we describe the numerical methods used
in our study. All of the algorithms were implemented
on a Silicon Graphics R4000 workstation.

We use two approaches to attack the problem
numerically: The first approach is to compute the exact
distribution $P_N(L,Q)$ for small values of $N$ by
considering all possible $N$-step walks. Since the
computational time increases exponentially with $N$,
this method practical only up to $N\approx 30$, and we have
analyzed  RWs with up to 36 steps this way. Thus,
in order to determine the scaling form $p(\ell,q)$,
it is necessary to use a random sampling of the set
$\Omega_N$ for large values of $N$. Using such a
Monte Carlo (MC) procedure, we have investigated
RWs of up to 1024 steps. Since the $q=0$ case is
especially interesting, we have used more efficient
algorithms to determine $p(l,0)$ to a higher accuracy.
For both the exact enumeration and MC calculations,
our algorithms require $O(N)$ operations to
process one sample from $\Omega_N$ for $p(l,0)$, and
$O(N^{3/2})$ operations to process the full probability
distribution $p(l,q)$. Further details on the individual
algorithms, as well as the algorithm used to
determine $p^{(d)}(\ell,{\bf 0})$ are given below.

\subsection{Algorithm for $p(\ell,0)$}
\label{alg1}

The only difference between exact enumeration and MC
algorithms involve the number of RWs analyzed:
In exact enumeration, the number of analyzed RWs
increases exponentially with $N$, whereas the
samples are chosen at random in the MC routines,
and the sample size is usually set to a constant.
Standard random number generators
are used to generate the RWs in the MC algorithm.
For each RW, the size of the largest loop is determined
and this is recorded in a histogram (with sizes from
0 to $L$) that eventually represents the probability
distribution we are looking for.
The determination of the largest neutral segment
in a given sequence is identical in both enumeration
and MC algorithms, and is described below.

Given a RW $\omega$, an array $F(Q)$ stores the step
number $i$ when $S_i(\omega)=Q$ for the first time.
Initially, $F(Q)=-1$ for all $Q$.
At each step of the RW (including step 0), the current step
number $i$ is recorded in $F(S_i(\omega))$ if
the site is visited for the first time, i.e. if
$F(S_i(\omega))=-1$.
If the site was visited earlier, the maximum loop size
is replaced by the maximum of itself and the difference
$i-F(S_i)$.
Since $F(S_i)$ stores the first time a site is visited,
the largest loop in the walk must correspond to one of such
differences. A finite number of operations are needed
for each step, therefore this part of the algorithm
involves $O(N)$ operations.

\subsection{Algorithm for $p(\ell,q)$}

The selection of RWs (enumeration or MC) and the creation of the
histogram are also straightforward for this more general
problem. The main task is to find an efficient algorithm
that produces the sizes of largest $Q$--segments (for all $Q$)
in a given sequence $\omega$. A straightforward generalization
of the algorithm for $p(\ell,0)$ would have required $O(N^2)$
operations per sequence. However, our algorithm takes advantage
of the fact that the same positions are visited many times,
and it requires only $O(N^{3/2})$ operations instead.
As usual, the algorithm traces the sequence one by one.
There are two main arrays. At a given step $i$, one of them
keeps track of
the sizes of largest $Q$--segments encountered that far.
The second array is actually a dynamically allocated list
of pairs of integers. Each pair in the list stores a
charge $q$ and size of the largest $q$--segment that ends
at the current step $i$. The size of this array grows
as $\sqrt{i}$ on the average. At each increment in
step size, all pairs in the list are updated by
adding the next element in the sequence to $q$ and
incrementing the corresponding lengths by one. These
lengths are then compared with the corresponding
values in the first array, which is updated if the
new length is larger. A new element is added to the
list of pairs whenever the walk reaches a position
for the first time, a condition that is checked for
separately. All the operations in an update can be
accomplished by a single pass through the list of
pairs, thus the whole algorithm requires only
$O(N^{3/2})$ operations to complete, as mentioned
earlier.

\subsection{Algorithm for $p^{(d)}(\ell,{\bf 0})$}

For the MC determination of $p^{(d)}(\ell,{\bf 0})$ at
higher dimensions, the $O(N)$ algorithm described
in Sec.~\ref{alg1} requires $O(N^d)$ storage elements
for the array $F({\bf Q})$, which quickly becomes
prohibitive with increasing $d$. The storage requirement
can be reduced to $O(dN)$ by storing the time series
of the position $S_i(\omega)$ of the RW instead.
However, the simplest algorithms would require $O(N^2)$
operations to find the largest {\bf 0}--segment given
such a data structure.
Note that the typical RW in dimensions $d\geq 2$ does not
revisit the same site more than a few times, and therefore
the total number of 0--segments in a RW should be only of $O(N)$.
We have taken advantage of this fact in order to
devise an algorithm that requires only $O(N\log N)$ operations
to do the job. The algorithm is as follows:

After the position array $S(i)$ is formed, its contents
[which are the position vectors $(Q_1,\cdots,Q_d)$]
are indexed in lexicographical order. This operation
requires only $O(N\log N)$ operations, when an efficient
sorting algorithm like Heapsort\cite{NR} is used.
All 0--segments in the sequence start and
end at the same position by definition, therefore the two
endpoints will be adjacent in the lexicographical index.
Going through the index sequentially, it is then possible
to determine the largest of the 0--segments in only $O(N)$
operations. The extraordinary speedup of this
algorithm makes is possible to go up to sample sizes of
$10^6$ for 1000--step RWs in 7 dimensions.

\end{appendix}

\begin{references}
\bibitem{proteins}See, e.g., T. E. Creighton, {\it Proteins:
Their Structure and Molecular Properties}, Freeman, San
Francisco (1984).
\bibitem{degen}P. G. de Gennes, {\it Scaling concepts in Polymer
Physics}, Cornell Univ. Press, Ithaca (1979).
\bibitem{HJ}P.G. Higgs and J.-F. Joanny, J. Chem. Phys. {\bf
94}, 1543 (1991).
\bibitem{KKrg}Y. Kantor and M. Kardar, Europhys. Lett. {\bf
14}, 421 (1991).
\bibitem{KKL}Y. Kantor, H. Li and M. Kardar, Phys. Rev. Lett.
{\bf 69}, 61 (1992). Y. Kantor, M. Kardar and H. Li, Phys. Rev.
{\bf E49}, 1383 (1994).
\bibitem{experiments}X.-H. Yu, A. Tanaka, K. Tanaka, and T.
Tanaka, J. Chem. Phys. {\bf 97}, 7805 (1992); X.-H. Yu, Ph.D.
thesis, MIT (1993); M. Annaka and T. Tanaka, Nature {\bf 355},
430 (1992); M. Scouri, J. P. Munch, S. F. Candau, S. Neyret, and
F. Candau, Macromol. {\bf 27}, 69 (1994).
\bibitem{KK}Y. Kantor and M. Kardar, Europhys. Lett. {\bf 27},
643  (1994), and Phys. Rev. {\bf E51}, 1299 (1995).
\bibitem{KKenum}Y. Kantor and M. Kardar, Phys. Rev. {\bf E52},
in press (1995).
\bibitem{rchandra}S. Chandrasekhar, Rev. Mod. Phys. {\bf 15},
1 (1943).
\bibitem{mathnote}W.~Feller, {\it An Introduction to Probability
Theory and Its Applications}, Vol.~1, 3rd Ed.,
John Wiley and Sons, New York (1968).
\bibitem{KEloop}Y. Kantor and D. Erta\c s, J. Phys. A: Math. Gen.
{\bf 27}, L907 (1994).
\bibitem{langfoot}In the language of RWs,
the term ``0--segment'' corresponds to a loop.  Throughout this
paper we will use  the ``language of
charged $N$-mers'' and the ``language of RWs'' interchangeably.
\bibitem{Feller2} Ref.~\protect\cite{mathnote}, p. 76.
\bibitem{NR}W.~H.~Press {\em et al.},
{\it Numerical Recipes in C: The Art of Scientific Computing},
2nd Ed., Cambridge Univ. Press, New York (1992), p. 336.
\end{references}
\end{document}